\documentclass[showpacs,twocolumn,preprintnumbers,amsmath,amssymb]{revtex4}
\usepackage{amsmath,amsfonts,latexsym,amssymb,graphicx,graphics,epsfig,subfigure,color,makeidx}
\usepackage{xcolor,diagbox}
\usepackage{multirow}
\usepackage[colorlinks,linkcolor=blue,anchorcolor=blue,citecolor=green,urlcolor=blue]{hyperref}
\usepackage{mathrsfs}
\usepackage{amsmath}

\begin{document}
\title{Sound from extra dimension: quasinormal modes of thick brane}

\author{Qin Tan$^{a}$$^{b}$\footnote{Qin Tan and Wen-Di Guo are co-first authors of the article.}}
\author{Wen-Di Guo$^{a}$$^{b}$}
\author{Yu-Xiao Liu$^{a}$$^{b}$\footnote{liuyx@lzu.edu.cn, corresponding author}}

\affiliation{$^{a}$Lanzhou Center for Theoretical Physics, Key Laboratory of Theoretical Physics of Gansu Province, School of Physical Science and Technology, Lanzhou University, Lanzhou 730000, China\\
$^{b}$Institute of Theoretical Physics and Research Center of Gravitation, Lanzhou University, Lanzhou 730000, China\\}
\begin{abstract}
In this work, we investigate the quasinormal modes of a thick brane system. Considering the transverse-traceless tensor perturbation of the brane metric, we obtain the Schr\"odinger-like equation of the Kaluza-Klein modes of the tensor perturbation. Then we use the  Wentzel-Kramers-Brillouin approximation and the asymptotic iteration method to solve this Schr\"odinger-like equation. We also study the numeric evolution of an initial wave packet against the thick brane. The results show that there is a set of discrete quasinormal modes in the thick brane model. These quasinormal modes appear as the decaying massive gravitons for a brane observer. They are characteristic modes of the thick brane and can reflect the structure of the thick brane.
\end{abstract}

\maketitle

\section{Introduction}
\label{Introduction}
As the characteristic modes of a dissipative system, quasinormal modes (QNMs) exist in every aspect of our world. These QNMs contain the key features that are characteristics of the physical systems. Studying them would help us to unravel the mysteries of the physical systems.  In black hole physics, the QNMs are thought to be able to carry information about black holes and have attracted much attention~\cite{Berti:2009kk,Kokkotas:1999bd,Nollert:1999ji,Konoplya:2011qq,Cardoso:2016rao,Jusufi:2020odz,Cheung:2021bol}, especially after the detection of gravitational waves~\cite{LIGOScientific:2016aoc}. Other physical systems such as leaky resonant cavities, QNMs also play an important role~\cite{Kristensen:2015qq}. So we are curious about what role QNMs might play in the braneworld model.

The braneworld models were originally introduced as a solution to the hierarchical problem between the weak and Plank scales. Among them, the warped extra dimension models proposed by Randall and Sundrum (RS) have attracted a lot of interest~\cite{Randall:1999ee,Randall:1999vf}. They consist one brane (RS-II model) or two branes (RS-I model) embedded in a five-dimensional anti-de-Sitter spacetime. Since the RS models were proposed, they have been studied in many realms such as particle physics, cosmology, and black hole physics. The applicability has gone far beyond its original scope~\cite{Shiromizu:1999wj,Tanaka:2002rb,Gregory:2008rf,Jaman:2018ucm,Adhikari:2020xcg,
Bhattacharya:2021jrn}. Combining the RS-II model~\cite{Randall:1999vf} and the domain wall model~\cite{Akama:1982jy,Rubakov:1983bb}, the thick brane models were developed~\cite{DeWolfe:1999cp,Gremm:1999pj,Csaki:2000fc}. It is a smooth extension of the RS-II model. The inclusion of brane thickness gives us new possibilities. Usually, most thick branes are generated by one or more scalar fields, but they can also be generated by a vector or spinor field~\cite{Dzhunushaliev:2010fqo,Dzhunushaliev:2011mm,Geng:2015kvs}. In order to recover the physics in our four-dimensional world, the zero modes of various fields should be confined on the brane. In previous literatures, some thick brane models and the localization of various matter fields on the brane were investigated~\cite{Melfo2006,Almeida2009,
Zhao2010,Chumbes2011,Liu2011,Xie2017,Gu2017,ZhongYuan2017,ZhongYuan2017b,Zhou2018,Chen:2020zzs,Hendi:2020qkk,Xie:2021ayr,Moreira:2021uod,Xu:2022ori,Silva:2022pfd,Xu:2022gth}. Besides these zero modes, there are massive Kaluza-Klein (KK) modes which might propagate along extra dimensions. These massive KK modes provide the possibility of detecting extra dimensions. In addition, cosmological thick brane solutions were also investigated~\cite{Mounaix:2002mm,Ghassemi:2006qk,Wu:2010stv}.

Quasinormal modes in higher dimensional theories also attract the interest of researchers~\cite{Chakraborty:2017qve,Dey:2020pth,Prasobh:2014zea,Hashemi:2019jlt,Chen:2016qii,Konoplya:2003dd,Cardoso:2003vt}. It is expected that the signatures of extra dimensions can be extracted from QNMs of black holes on the brane. These signals can be used to constrain the extra dimensional models~\cite{Seahra:2004fg,Seahra:2006tm,Chung:2015mna,Dey:2020lhq,Banerjee:2021aln,Mishra:2021waw,Lin:2022hus}. But these researches are mainly focused on the QNMs of black holes on the brane. Does a brane have a characteristic sound? That is, does it have a set of discrete QNMs as characteristic modes of a braneworld model? For the RS-II model, the answer is yes~\cite{Seahra:2005wk,Seahra:2005iq}. Seahra studied the scattering of KK gravitons in the RS-II model and found that the brane possesses a series of discrete QNMs~\cite{Seahra:2005wk,Seahra:2005iq}.

Reference~\cite{Tan:2022uex} investigated the evolution of massive modes in the thick brane model and found that the evolution behavior is similar to QNMs. This arouse our interest in QNMs in thick brane models. As far as we know, the QNMs of a thick brane have not been investigated. As the characteristic modes of a brane, it can reflect the structure of the thick brane. On the other hand, since the QNMs dominating the time evolution of some initial fluctuations from the physic system's equilibrium state, they can be used to verify the stability of the brane~\cite{Clarkson:2005mg}. It is undoubtedly interesting to study the QNMs of a thick brane. We will use semi-analytical and numerical methods to study QNMs in a thick brane model.

The organization of the rest of this paper is as follows. In Sec.~\ref{BRANE WORLD MODEL}, we review a solution of the thick brane and the linear metric tensor perturbation. Based on this solution, we solve for the QNMs of this thick brane. In Sec.~\ref{Quasi-normal modes of thick brane}, we compute the quasinormal frequencies of the thick brane by using semi-analytical methods. We also compare them with the results of numerical evolution. Finally, Sec.~\ref{Conclusion} gives the conclusions and discussions.

\section{Braneworld model in general relativity}
\label{BRANE WORLD MODEL}
In this section, we will briefly review the thick brane solution and its gravitational perturbation. Usually, a thick brane can be generated by a wide variety of matter fields like scalar fields and vector fields. Here we choose a canonical scalar field to generate the brane. The action of this thick brane model is the Einstein-Hilbert action minimally coupled to a canonical scalar field
\begin{eqnarray}
	S=\int d^5x\sqrt{-g}\left(\frac{1}{2\kappa^{2}_{5}}R-\frac{1}{2}g^{MN}\partial_{M}
	\varphi\partial_{N}\varphi -V(\varphi)\right),\label{action}
\end{eqnarray}
where $\kappa_{5}$ is the five-dimensional gravitational constant. Hereafter, capital Latin letters $M,N,\dots=0,1,2,3,5$ label the five-dimensional indices, while Greek letters $\mu,\nu\dots=0,1,2,3$ and Latin letters $i,j\dots=1,2,3$ label the four-dimensional ones and three-dimensional space ones on the brane, respectively. The dynamical field equations are
\begin{eqnarray}
R_{MN}-\frac{1}{2}Rg_{MN}&=&-g_{MN}\kappa^{2}_{5}\left(\frac{1}{2}\partial^{A}\varphi\partial_{A}\varphi
+V(\varphi)\right)\nonumber\\
&&+\kappa^{2}_{5}\partial_{M}\varphi\partial_{N}\varphi,\label{field equation}\\
g^{MN}\nabla_{M}\nabla_{N}\varphi&=&\frac{\partial V(\varphi)}{\partial\varphi}.\label{motion equation}
\end{eqnarray}
The five-dimensional metric ensuring the four-dimensional Poincar\'{e} symmetry is~\cite{Csaki:2000fc}
\begin{equation}
	ds^2=e^{2A(y)}\eta_{\mu\nu}dx^\mu dx^\nu+dy^2,
	\label{metric}
\end{equation}
where $\eta_{\mu\nu}=\text{diag}(-1,1,1,1)$ is the four-dimensional Minkowski metric. Now, the specific dynamical equations can be written as
\begin{eqnarray}
	6A'^2 +3A''&=&-\frac{\kappa^{2}_{5}}{2}\varphi'^2-\kappa^{2}_{5}V,  \label{EoMs1}\\
	6A'^2&=&\frac{\kappa^{2}_{5}}{2}\varphi'^2-\kappa^{2}_{5}V,  \label{EoMs2}\\
	\varphi{''}+4A'\varphi'&=&\frac{\partial V}{\partial\varphi},  \label{EoMsphi}
\end{eqnarray}
where prime denotes the derivative with respect to the extra dimensional coordinate $y$. By using the first-order formalism, the thick brane solution was investigated in Ref.~\cite{Gremm:1999pj}:
\begin{eqnarray}
A(y)&=&-b\ln\left(\cosh(ky)\right),\label{warpfactorsolution1}\\
\varphi(y)&=&\sqrt{\frac{3b}{\kappa^2_{5}}}\arcsin\left(\tanh\left(ky\right)\right),\label{scalarfieldsolution1}\\
V(\varphi)&=&\frac{3bk^{2}}{4\kappa^{2}_{5}}\left(1-4b+(1+4b)\cos\left(\sqrt{\frac{4\kappa^{2}_{5}}{3b}}\varphi\right)\right).\nonumber\\\label{scalarpotentialsolution1}
\end{eqnarray}
Here, $b$ is a dimensionless parameter and $k$ is a parameter with mass dimension one. If we choose $\kappa_{5}=\sqrt{2}$, the above solutions for the scalar field $\varphi$ and potential $V$ are the same to Ref.~\cite{Gremm:1999pj} because $\arcsin\left(\tanh\left(ky\right)\right)=2\arctan\left(\tanh\left(ky/2\right)\right)$\footnote{There is a typo in the scalar potential $V$ in Ref.~\cite{Gremm:1999pj}, the $-$ in front of $(1+4b)$ should be replaced by $+$.}. Besides, the warp factor (\ref{warpfactorsolution1}) differs from the one in Ref.~\cite{Gremm:1999pj} a factor $2$ in front of $\cosh(ky)$. It does not matter, because this factor can be absorbed into a new four-dimensional coordinate. Next, we consider the linear transverse-traceless tensor perturbation of the metric. The perturbed metric is given by
\begin{eqnarray}
	g_{MN}=\left(
	\begin{array}{cc}
		e^{2A(y)}(\eta_{\mu\nu}+h_{\mu\nu}) & 0\\
		0 & 1\\
	\end{array}
	\right)\label{perturbed metric},
\end{eqnarray}
where $h_{\mu\nu}$ satisfies the transverse-traceless condition
\begin{eqnarray}
\partial_{\mu}h^{\mu\nu}=0=\eta^{\mu\nu}h_{\mu\nu}.\label{TTguage}
\end{eqnarray}
Substituting the perturbed metric \eqref{perturbed metric} into the field equation \eqref{field equation}, we obtain the linear equation of the tensor fluctuation:
\begin{eqnarray}
\left(e^{-2A}\Box^{(4)}h_{\mu\nu}+h''_{\mu\nu}+4A'h'_{\mu\nu}\right)=0, \label{mainequation}
\end{eqnarray}
where $\Box^{(4)}=\eta^{\alpha\beta}\partial_{\alpha}\partial_{\beta}$. Introducing the following coordinate transformation $dz=e^{-A}dy$, the metric~(\ref{metric}) can be written as
\begin{equation}
	ds^2=e^{2A(z)}(\eta_{\mu\nu}dx^\mu dx^\nu+dz^2),
	\label{conformalmetric}
\end{equation}
and Eq.~(\ref{mainequation}) becomes
\begin{equation}
\left[\partial^{2}_{z}+3(\partial_{z}A)\partial_{z}+\Box^{(4)}\right]h_{\mu\nu}=0.\label{conformalequation1}
\end{equation}
The perturbation $h_{\mu\nu}$ can be written as \cite{Seahra:2005iq}
\begin{equation}
h_{\mu\nu}=e^{-\frac{3}{2}A(z)}\Phi(t,z)e^{-i a_{j}x^{j}}\epsilon_{\mu\nu}, ~~~~\epsilon_{\mu\nu}=\text{constant}.\label{decomposition1}
\end{equation}
Substituting the above decomposition (\ref{decomposition1}) into Eq.~(\ref{conformalequation1}), we obtain a one-dimensional wave equation of $\Phi(t,z)$
\begin{equation}
	-\partial_{t}^{2}\Phi+\partial_{z}^{2}\Phi-U(z)\Phi-a^{2}\Phi=0, \label{evolutionequation}
\end{equation}
where
\begin{eqnarray}
U(z)=\frac{3}{2}\partial_{z}^{2}A+\frac{9}{4}(\partial_{z}A)^{2} \label{effectivepotential}
\end{eqnarray}
is the effective potential and $a$ is a constant coming from the separation of variables. Furthermore, separability means that the function $\Phi(t,z)$ can be decomposed as
\begin{eqnarray}
\Phi(t,z)=e^{-i\omega t}\phi(z).\label{decomposition2}
\end{eqnarray}
So we can obtain a Schr\"odinger-like equation of the extra dimensional part $\phi(z)$
\begin{equation}
	-\partial_{z}^{2}\phi(z)+U(z)\phi(z)=m^{2}\phi(z),\label{Schrodingerlikeequation}
\end{equation}
where $m^2=\omega^2-a^2$ is the mass of the KK modes. Equation \eqref{Schrodingerlikeequation} supports a bound zero mode $\phi_{0}(z) \propto e^{\frac{3}{2}A(z)}$ which is localized on the brane for the solution (\ref{warpfactorsolution1}) with $b>0$, and a series of massive KK modes. Usually, the massive KK modes stay on the brane for a finite time and eventually escape to infinity of the extra dimension. Thus the thick brane is a dissipative system for the massive KK modes. Similar to QNMs in the black hole system, there are also characteristic modes with complex frequencies in the thick brane model. These modes can also reflect the properties of the thick brane model. We will discuss them in the next section.

\section{Quasinormal modes of thick brane}
\label{Quasi-normal modes of thick brane}
\begin{figure}
	\subfigure[~The effective potential~(\ref{effpotentialform})]{\label{figeffective1}
		\includegraphics[width=0.22\textwidth]{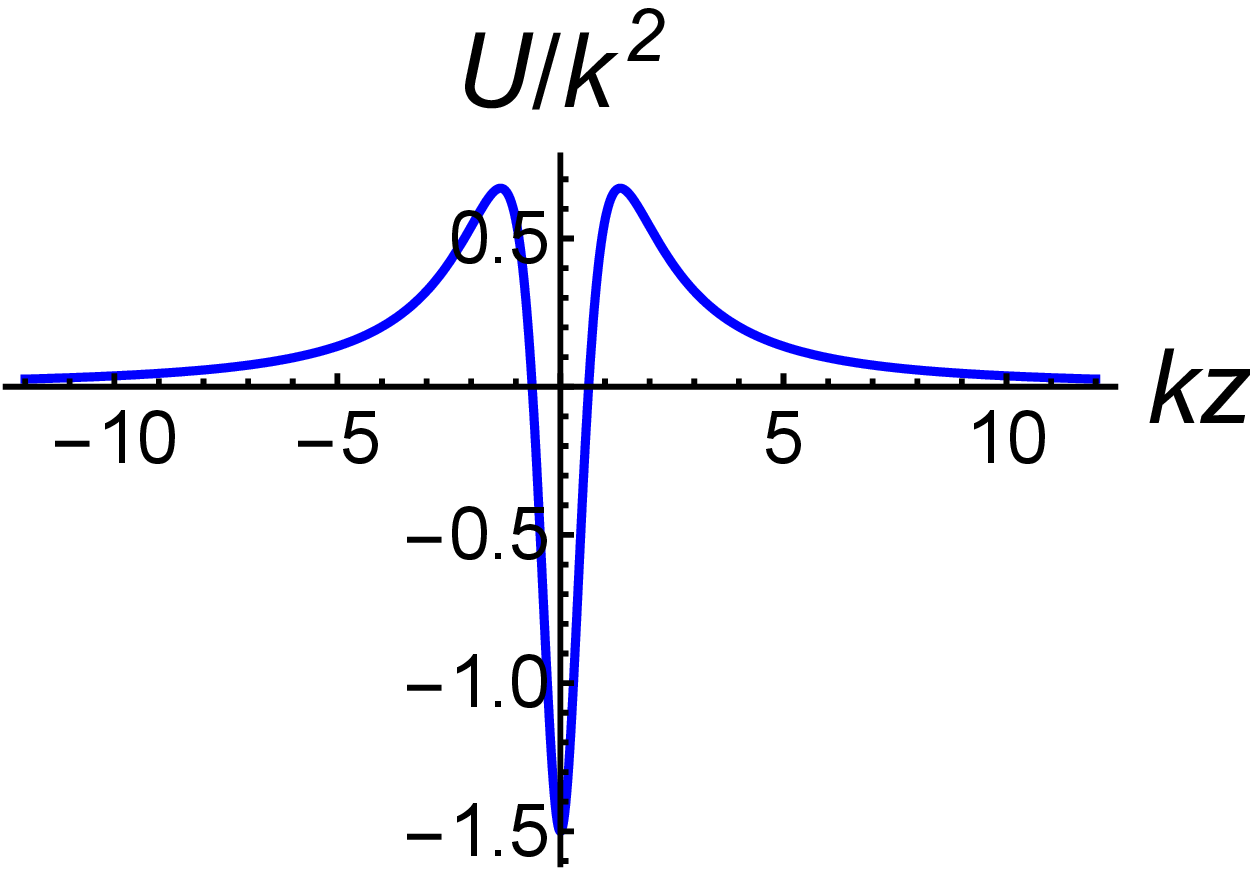}}
		\subfigure[~The dual effective potential~(\ref{dualeffpotentialform})]{\label{figeffective2}
		\includegraphics[width=0.22\textwidth]{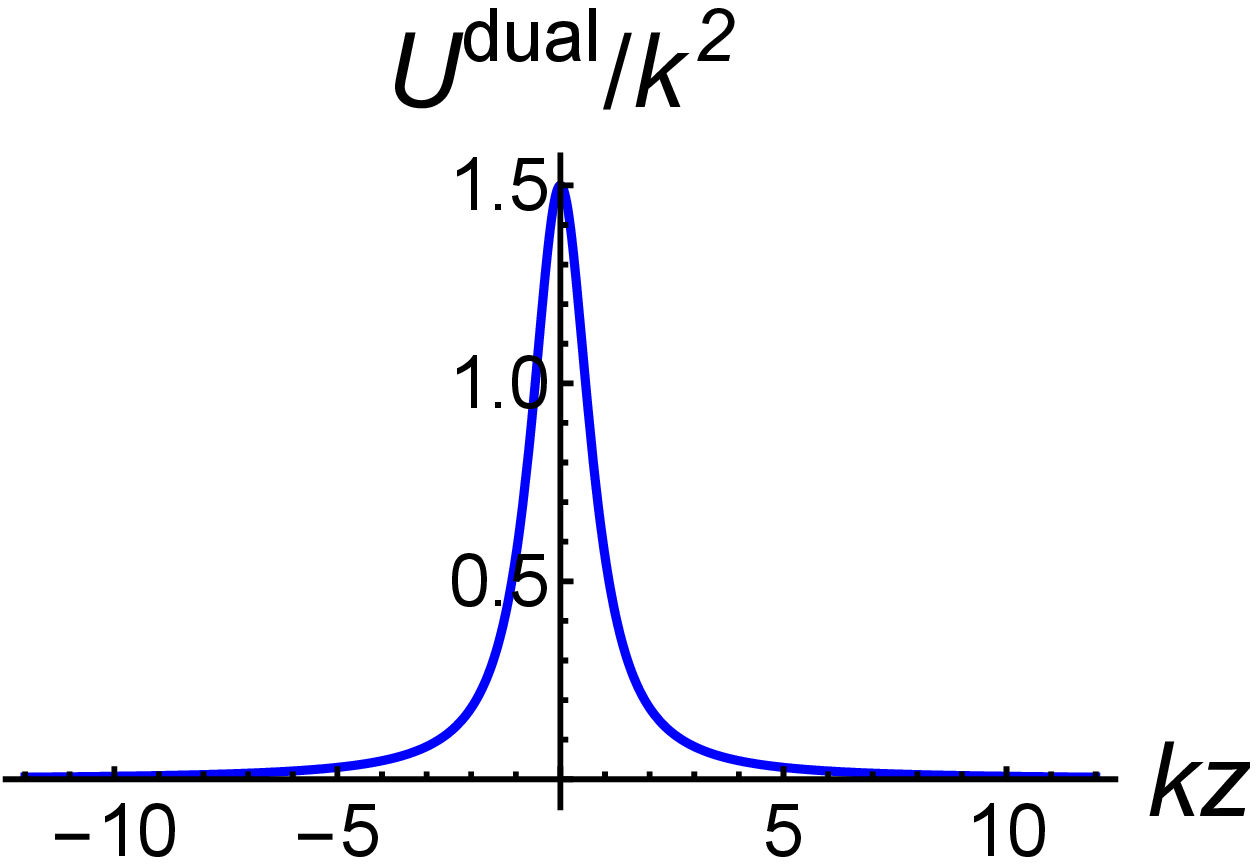}}
	\subfigure[~The zero mode~(\ref{zeromode})]{\label{figzeromode}
	\includegraphics[width=0.22\textwidth]{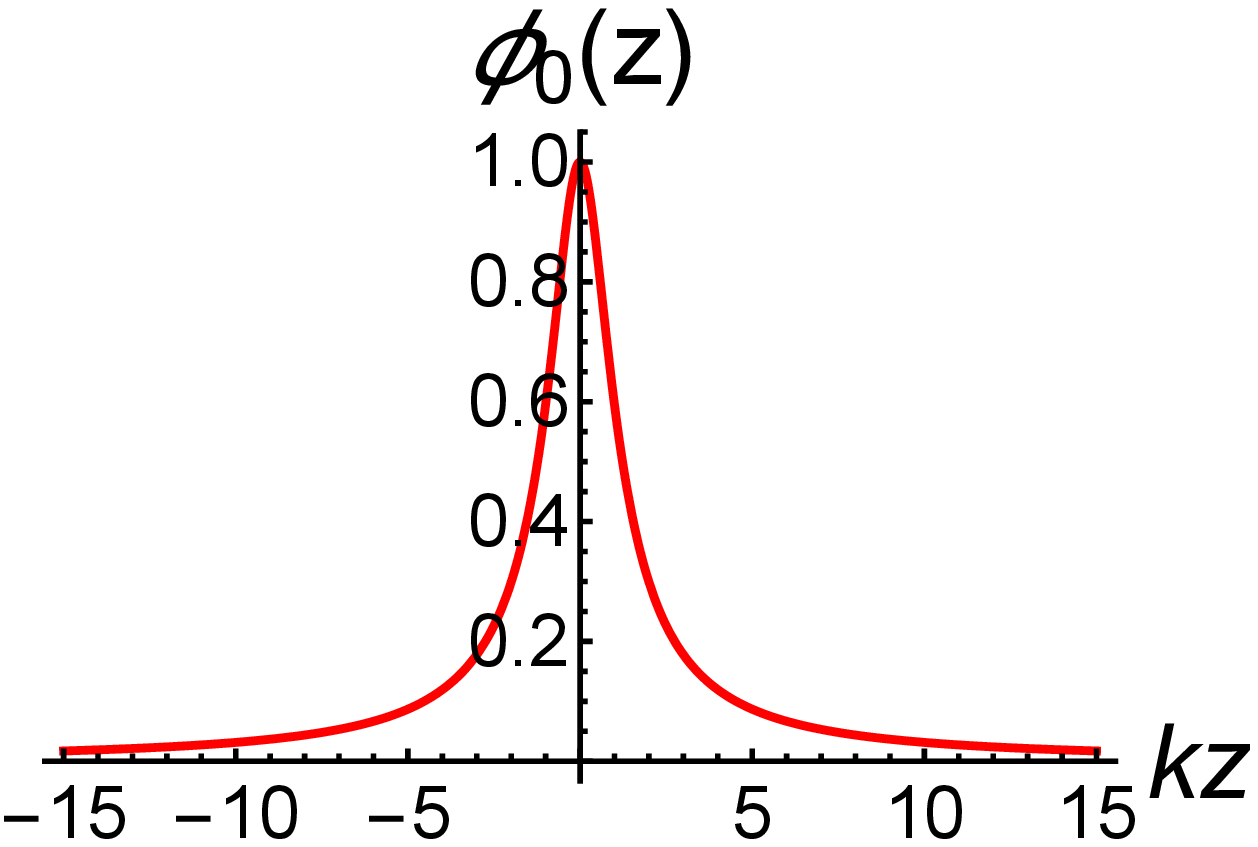}}	
	\caption{The shapes of the effective potential~(\ref{effpotentialform}), the dual effective potential~(\ref{dualeffpotentialform}), and the zero mode \eqref{zeromode}.}\label{0modeandU}
\end{figure}

In this section, we will use some semi-analytical methods to solve the QNMs of the thick brane. As can be seen from the Schr\"odinger-like equation~(\ref{Schrodingerlikeequation}), it is the effective potential $U(z)$ that determines the QNMs. Substituting the thick brane solution~(\ref{warpfactorsolution1}) into the effective potential~(\ref{effectivepotential}), we can obtain the specific form of the effective potential. Note that we only consider $b=1$, because the coordinate transformation relation between $y$ and $z$ is analytical for this case. The specific forms of the warp factor $A(z)$, the effective potential $U(z)$, and the zero mode $\phi_{0}(z)$ are given by
\begin{eqnarray}
A(z)&=&-\frac{1}{2}\ln(k^{2}z^{2}+1),\label{zcoorwarpfactor}\\
U(z)&=&\frac{3 k^2 \left(5 k^2 z^2-2\right)}{4 \left(k^2 z^2+1\right)^2},\\\label{effpotentialform}
\phi_{0}(z)&=&\frac{1}{(1+k^{2}z^{2})^{3/4}}.\label{zeromode}
\end{eqnarray}
We plot the effective potential and the zero mode in Fig.~\ref{0modeandU}. It can be seen that the effective potential is volcano-like and $U(z)\rightarrow0$ when $|z|\rightarrow\infty$. This potential is a smooth extension of the effective potential in the RS-II model. In the thin brane scenario, the general solution of the massive KK modes is the Hankel function. So the QNMs can be analytically obtained by imposing the outgoing boundary condition to the Hankel function~\cite{Seahra:2005wk}. But there is no any analytical solution of massive KK modes for this thick brane. So we use some semi-analytical method to obtain the QNMs of the thick brane. Unlike the case of a black hole, there is a potential well but not a pure barrier for our brane case. Therefore, some methods of solving the QNMs commonly used in black holes, such as the Wentzel-Kramers-Brillouin (WKB) approximation~\cite{Konoplya:2003ii}, cannot solve the QNMs of this thick brane directly. But we notice that the Schr\"odinger-like equation~(\ref{Schrodingerlikeequation}) can be factorized as a super-symmetric form
\begin{equation}
QQ^{\dagger}\phi(z)=m^{2}\phi(z)\label{supersymmetricform},
\end{equation}
where $Q$ and $Q^{\dagger}$ are
\begin{equation}
Q=\partial_{z}+\frac{3}{2}\partial_{z}A	,~~~~~~~~Q^{\dagger}=-\partial_{z}+\frac{3}{2}\partial_{z}A.
\end{equation}
The above equation~(\ref{supersymmetricform}) has a corresponding Schr\"odinger-like equation with the super-symmetric partner potential:
\begin{eqnarray}
Q^{\dagger}Q\tilde{\phi}(z)=\left(-\partial_{z}^{2}+U^{dual}(z)\right)\tilde{\phi}(z)=m^{2}\tilde{\phi}(z),\label{dualSchrodingerlikeequation11}
\end{eqnarray}
where
\begin{eqnarray}
	U^{\text{dual}}(z)=-\frac{3}{2}\partial_{z}^{2}A+\frac{9}{4}(\partial_{z}A)^{2}=\frac{3 k^2 \left(k^2 z^2+2\right)}{4 \left(k^2 z^2+1\right)^2}.\label{dualeffpotentialform}
\end{eqnarray}
According to the super-symmetric quantum mechanics, the Schr\"odinger-like equations~(\ref{Schrodingerlikeequation}) and \eqref{dualSchrodingerlikeequation11} have the same spectrum of massive KK modes~\cite{Cooper:1994eh}. So the effective potential and the super-symmetric partner potential have the same spectrum of QNMs~\cite{Ge:2018vjq}. Plot of the super-symmetric partner potential \eqref{dualeffpotentialform} is  shown in Fig.~\ref{figeffective2}. We can see that the shape of the dual potential is similar to the effective potentials in the case of the Schwarzschild black hole, for which the QNMs can be solved by the asymptotic iteration method~\cite{Ciftci:2003As,ciftci:2005co,Cho:2011sf} and the WKB approximation. Therefore, we can obtain the quasinormal frequencies of the thick brane by using the dual potential~\eqref{dualeffpotentialform}.
 \subsection{Solve the QNMs of thick brane by using the asymptotic iteration method}
First, we use the asymptotic iteration method to solve the QNMs of the thick brane. Then we compare the results with those obtained by the WKB approximation. At the beginning, we give a brief review on the idea of the asymptotic iteration method. Consider a second-order homogeneous linear differential equation for the function $y(x)$
\begin{equation}
y''(x)=\lambda_{0}(x)y'(x)+s_{0}(x)y(x),\label{2orderdiffeq}
\end{equation}
where $\lambda_{0}(x)\neq0$ and $s_{0}(x)$ are $C^{\infty}$ functions. Based on the symmetric structure of the right-hand side of Eq.~\eqref{2orderdiffeq}, a general solution can be solved. Indeed, differentiating Eq.~(\ref{2orderdiffeq}) with respect to $x$, we find that
\begin{equation}
	y'''(x)=\lambda_{1}(x)y'(x)+s_{1}(x)y(x),
\end{equation}
where
\begin{eqnarray}
\lambda_{1}(x)&=&\lambda'_{0}+s_{0}+\lambda_{0}^{2},\\
s_{1}(x)&=&s'_{0}+s_{0}\lambda_{0}.
\end{eqnarray}
Iteratively, the $(n-1)$-th and $n$-th differentiations of Eq.~(\ref{2orderdiffeq}) give
\begin{eqnarray}
y^{(n+1)}(x)&=&\lambda_{n-1}(x)y'(x)+s_{n-1}(x)y(x),\\
y^{(n+2)}(x)&=&\lambda_{n}(x)y'(x)+s_{n}(x)y(x),	
\end{eqnarray}
where
\begin{eqnarray}
\lambda_{n}(x)&=&\lambda'_{n-1}+s_{n-1}+\lambda_{0}\lambda_{n-1}, \label{AIMrelation1}\\ s_{n}(x)&=&s'_{n-1}+s_{0}\lambda_{n-1}\label{AIMrelation2}.
\end{eqnarray}
The asymptotic aspect is introduced as follows for sufficiently large $n$
\begin{eqnarray}
\frac{s_{n}(x)}{\lambda_{n}(x)}=\frac{s_{n-1}(x)}{\lambda_{n-1}(x)}=\beta(x).\label{QNMscondition1}
\end{eqnarray}
We can obtain the QNMs from the ``quantization condition"
\begin{eqnarray}
s_{n}(x)\lambda_{n-1}(x)-s_{n-1}(x)\lambda_{n}(x)=0.\label{QNMscondition2}
\end{eqnarray}
To be more precise, we adopt the improved version of the asymptotic iteration method by Cho $et~al.$~\cite{Cho:2011sf}. The original asymptotic iteration method has the ``weakness'' that for each iteration one must take the derivative of the $s(x)$ and $\lambda(x)$ terms of the previous iteration. This might bring difficulties for numerical calculations. Cho $et~al.$ reduced the asymptotic iteration method into a set of recursion relations which no longer require derivative operators. This greatly improves the speed and precision of numerical calculation. In the asymptotic iteration method, when solving Eq.~\eqref{QNMscondition2}, we should take a specific point $\chi$. The two functions $\lambda_{n}$ and $s_{n}$ can be expanded in a Taylor series at the point $\chi$:
\begin{eqnarray}
\lambda_{n}(x)&=&\sum_{i=0}^{\infty}c_{n}^{i}(x-\chi)^{i},\\
s_{n}(x)&=&\sum_{i=0}^{\infty}d_{n}^{i}(x-\chi)^{i}.
\end{eqnarray}
Here, $c_{n}^{i}$ and $d_{n}^{i}$ denote the $i$-th Taylor coefficients of $\lambda_{n}$ and $s_{n}$, respectively. Substituting the above expressions into Eqs.~\eqref{AIMrelation1} and ~\eqref{AIMrelation2}, we can obtain a set of recursion relations
\begin{eqnarray}
c_{n}^{i}&=&(i+1)c_{n-1}^{i+1}+d^{i}_{n-1}+\sum_{k=0}^{i}c_{0}^{k}c_{n-1}^{i-k},\\
d_{n}^{i}&=&(i+1)d_{n-1}^{i+1}+\sum_{k=0}^{i}d_{0}^{k}c_{n-1}^{i-k}.
\end{eqnarray}
Now the ``quantization condition''~\eqref{QNMscondition2} can be rewritten as
\begin{equation}
d_{n}^{0}c_{n-1}^{0}-d_{n-1}^{0}c_{n}^{0}=0\label{QNMscondition3}.
\end{equation}
In this way, the ``quantization condition''~\eqref{QNMscondition2} reduced to a set of recursion relations which do not require derivative operators.

\begin{figure}[htbp]
	\centering
	\includegraphics[width=0.4\textwidth]{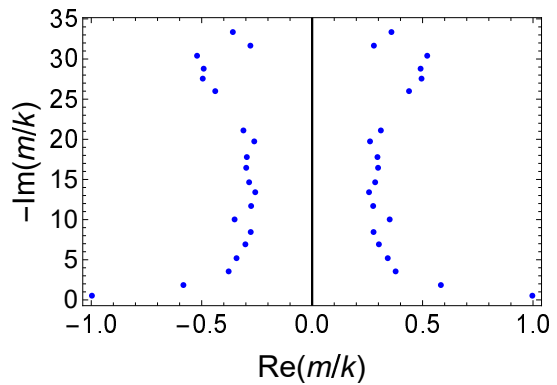}
	\caption{The first twenty quasinormal frequencies of the thick brane solved by the asymptotic iteration method. The iteration of asymptotic iteration method is 150.}\label{qnmsplot}
\end{figure}

The Schr\"odinger-like equation with the dual potential is
\begin{eqnarray}
-\partial_{z}^{2}\tilde{\phi}(z)+\left(\frac{3 k^2 \left(k^2 z^2+2\right)}{4 \left(k^2 z^2+1\right)^2}-m^{2}\right)\tilde{\phi}(z)=0.\label{dualSchrodingerlikeequation}
\end{eqnarray}
The boundary conditions are
\begin{equation}
	\label{boundaryconditions}
\tilde{\phi}(z) \propto \left\{
\begin{aligned}
e^{im z}, &~~~~~z\to\infty.& \\
e^{-im z},  &~~~~~z\to-\infty.&
\end{aligned}
\right.
\end{equation}
Obviously, there is no first derivative term in the above equation, which means $\lambda_{0}=0$. The asymptotic iteration method cannot be used directly in this situation. We need to transform our coordinates to obtain the equation whose first derivative term is nonvanishing. On the other hand, transforming the infinity to be finite is necessary. So we perform the transformation $u= \frac{\sqrt{4k^2 z^2+1}-1}{2k z}$. Then, Eq.~\eqref{dualSchrodingerlikeequation} becomes
\begin{eqnarray}
\frac{\left(u^2-1\right)^3 \left(\left(u^4-1\right) \tilde{\phi} ''(u)+2 u \left(u^2+3\right) \tilde{\phi}
	'(u)\right)}{\left(u^2+1\right)^3}\nonumber\\
+\left(\frac{m^2}{k^2}-\frac{3 \left(u^2-1\right)^2 \left(2 u^4-3 u^2+2\right)}{4
	\left(u^4-u^2+1\right)^2}\right) \tilde{\phi} (u)=0,\label{dualSchrodingerlikeequation1}
\end{eqnarray}
where $-1<u<1$. The boundary conditions~(\ref{boundaryconditions}) can be rewritten as
\begin{equation}
	\label{transformboundaryconditions}
	\tilde{\phi}(u) \propto \left\{
	\begin{aligned}
e^{-\frac{i m/k }{2 u-2}}, &~~~ u\to 1.& \\
e^{\frac{i m/k }{2 u+2}}, &~~~ u\to -1.&
	\end{aligned}
	\right.
\end{equation}
Thus, $\tilde{\phi}(u)$ can be written in the form
\begin{eqnarray}
	\tilde{\phi}(u)=\psi (u) e^{-\frac{i m/k }{2 u-2}} e^{\frac{i m/k }{2 u+2}}.\label{boundarysolutions}
\end{eqnarray}
Now the boundary condition becomes that the function $\psi(u)$ is finite at $u \to \pm 1$.
Substituting the expression~(\ref{boundarysolutions}) into Eq.~(\ref{dualSchrodingerlikeequation1}), we have
\begin{equation}
	\psi''(u)=\lambda_{0}(u)\psi'(u)+s_{0}(u)\psi(u),\label{dualSchrodingerlikeequation2}
\end{equation}
where
\begin{eqnarray}
\lambda_{0}(u)&=&-\frac{2 u \left(u^4+2 i \left(u^2+1\right) \frac{m}{k} +2 u^2-3\right)}{\left(u^2-1\right)^2 \left(u^2+1\right)},\label{lambda0}\\
s_{0}(u)&=&\frac{1}{4 \left(u^2+1\right) \left(u^6-2 u^4+2 u^2-1\right)^2}\nonumber\\
&&\times \Bigg[-4 \left(u^4-u^2+1\right)^2 \left(u^2+1\right)\frac{m ^2}{k ^2}\nonumber\\
&&+8 i \left(u^2-1\right) \left(u^4-u^2+1\right)^2 \frac{m}{k} \nonumber\\
&&+3 \left(2 u^4-3 u^2+2\right) \left(u^2+1\right)^3\Bigg].\label{s0}
\end{eqnarray}
With $\lambda_{0}$ and $s_{0}$ obtained, we can solve the quasinormal frequencies of the thick brane using the reduced ``quantization condition''~(\ref{QNMscondition3}). Using this method we obtain several QNMs of the thick brane. Plot of the first twenty QNMs obtained by the asymptotic iteration method is shown in Fig.~\ref{qnmsplot}. It can be seen that all the QNMs obtained by the asymptotic iteration method have a negative imaginary part. This means that the QNMs will dissipate.

We also compute the quasinormal frequencies through the WKB approximation~\cite{Konoplya:2019hlu}. In black hole physics, the WKB method was first applied to the scattering problem around black holes by Schutz and Will~\cite{schutz:1985bl}. The method is based on matching of the asymptotic WKB solutions at the event horizon and spatial infinity with the Taylor expansion near the peak of the potential barrier through the two turning points. Since the shape of the dual potential in the thick brane is similar to the effective potential in the case of the Schwarzschild black hole, the QNMs of the thick brane can be solved by the WKB approximation. Here we use the sixth order WKB approximation to solve the QNMs of the thick brane. The form of the sixth order WKB formula is
\begin{eqnarray}
i\frac{\omega^{2}-U_{\text{max}}}{\sqrt{-2U_{\text{max}}''}}-\sum_{j=2}^{6}\Lambda_{j}=n+1/2,~~~n=1,2,3... , \label{sixthoderwkb}
\end{eqnarray}
where $U_{\text{max}}$ is the maximum value of the dual potential, $\Lambda_{j}$ is the correction term of the $j$-th order that depends on the value of the dual potential and its derivatives at the peak value. The explicit form of the correction term $\Lambda_{j}$ can be found in Refs.~\cite{Iyer:1986np,Konoplya:2004ip,Konoplya:2003dd}. We can solve the QNMs of the thick brane using the above expression. The results are listed in Table~\ref{tab1}. Since the WKB approximation is more applicable to low overtones, i.e., QNMs with a small imaginary part. When the overtone number $n$ is moderately higher, the results of the WKB approximation become discredited~\cite{Berti:2009kk}. Therefore, we neglect $n\geq4$ for the results of the WKB approximation. We can see that for the first three QNMs, the results of the asymptotic iteration method are in good agreement with the results of the WKB approximation. This increases the credibility of our results. For higher overtone modes, we expect to explore new methods to compare with the results of the asymptotic iteration method.

\begin{table}[htbp]
\begin{tabular}{|c|c|c|}
\hline
$\;\;n\;\;$  &
$\;\;\text{Asymptotic iteration method}\;\;$  &
$\;\;\;\;\;\;\;\;\text{WKB method}\;\;\;\;\;\;\;$ \\
\hline
~   &~~~~$\text{Re}(m/k)$  ~~  $\text{Im}(m/k)~~$  &$~~~~~~\text{Re}(m/k)$ ~~ $\text{Im}(m/k)~~$       \\
1   &~~0.997018~~ -0.526362       &~~~1.04357~~~  -0.459859       \\
2   &0.581489~~ -1.85128      &~~0.536087~~ -1.71224  \\
3   &0.306005~~ -3.53366        &~~0.279715~~ -3.70181   \\
\hline
\end{tabular}
\caption{Low overtone modes using the asymptotic iteration method and WKB method.\label{tab1}}
\end{table}

\subsection{Evolution of initial wave packet}
\begin{figure}
	\subfigure[~$kz_{\text{ext}}=0$]{\label{figevenpofilezero0}
		\includegraphics[width=0.22\textwidth]{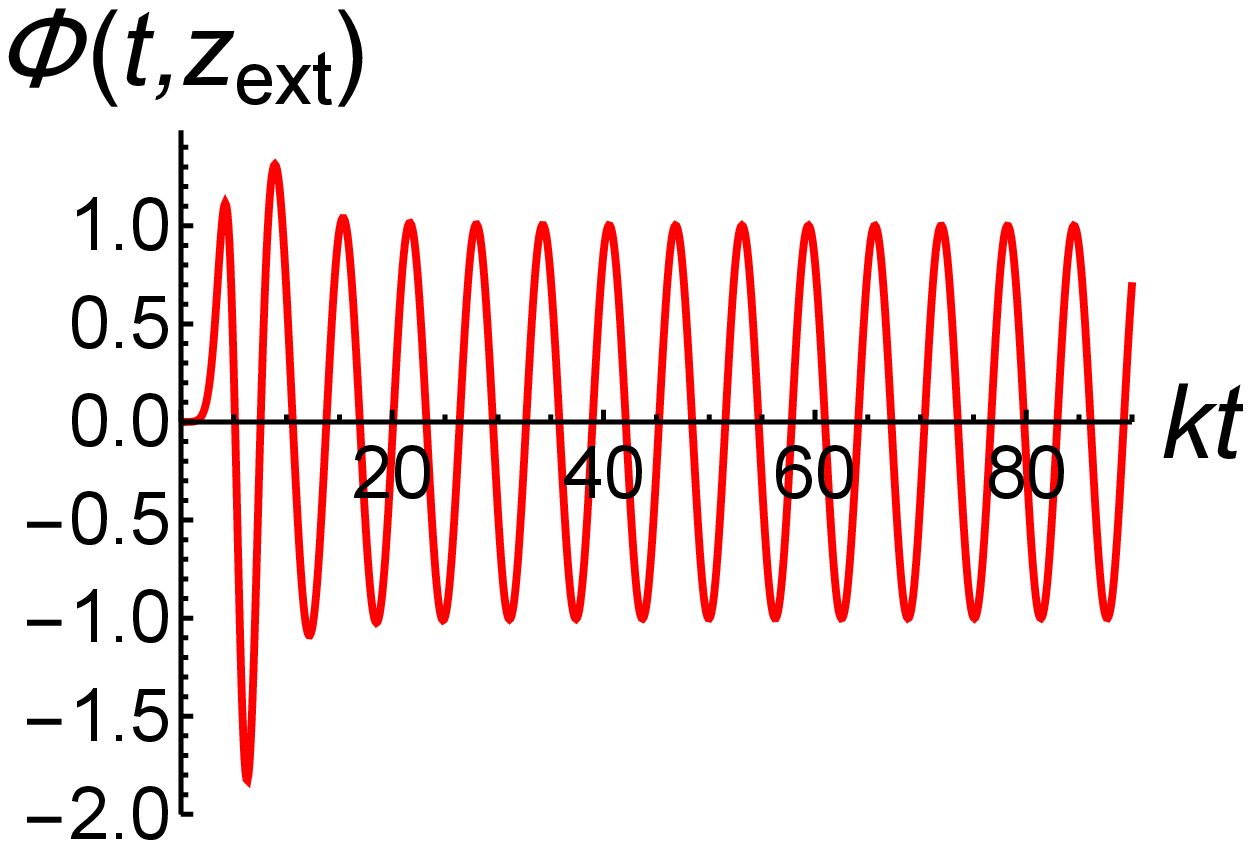}}
	\subfigure[~$kz_{\text{ext}}=0$]{\label{figlogevenpofilezero0}
		\includegraphics[width=0.22\textwidth]{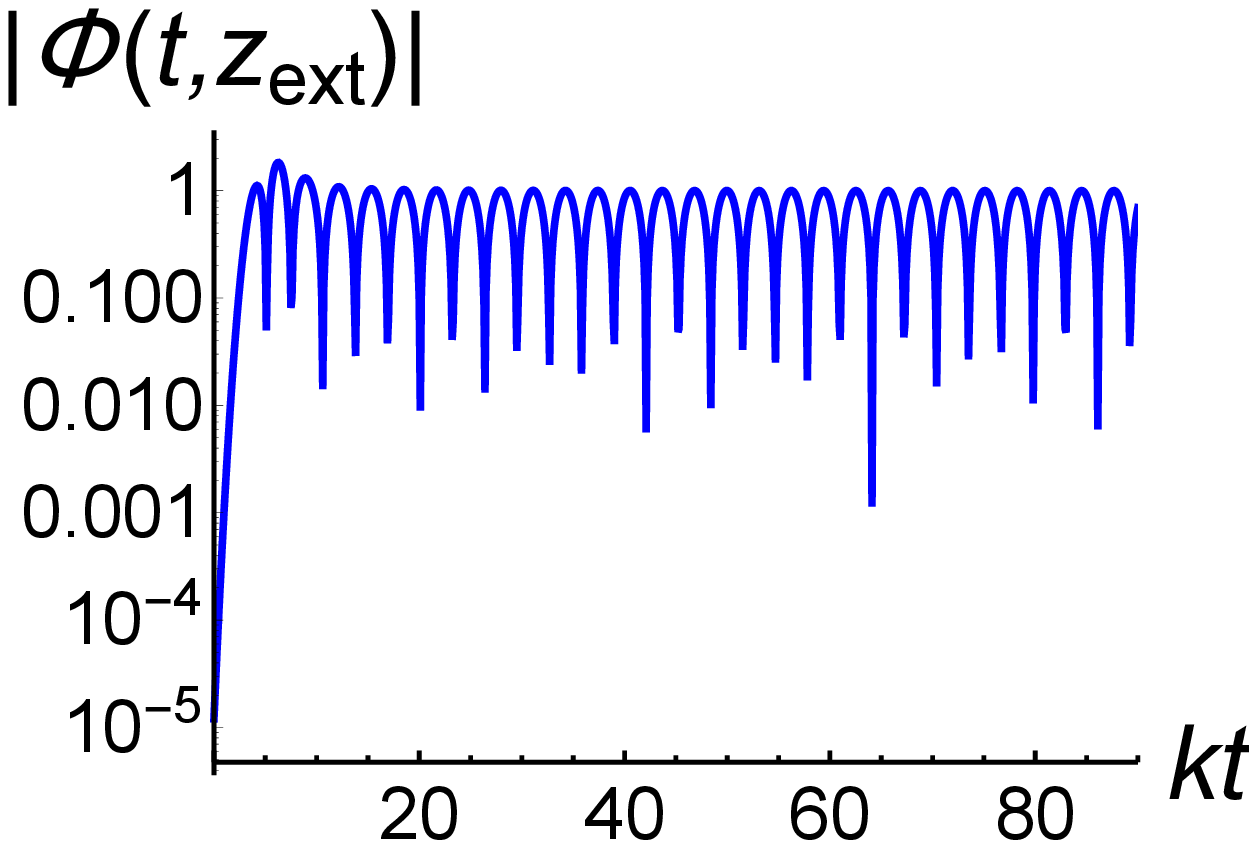}}
		\subfigure[~$kz_{\text{ext}}=3$]{\label{figevenpofilezero3}
		\includegraphics[width=0.22\textwidth]{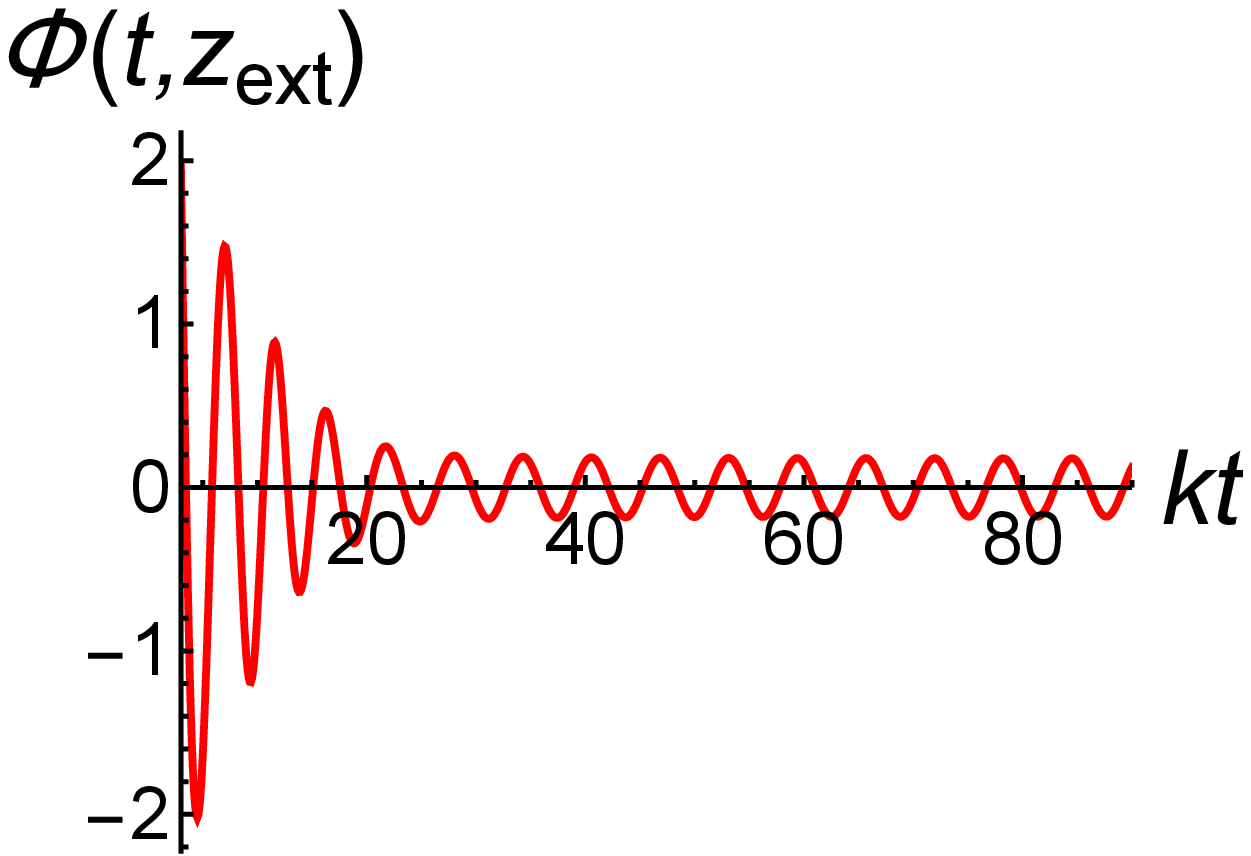}}
	\subfigure[~$kz_{\text{ext}}=3$]{\label{figlogevenpofilezero3}
		\includegraphics[width=0.22\textwidth]{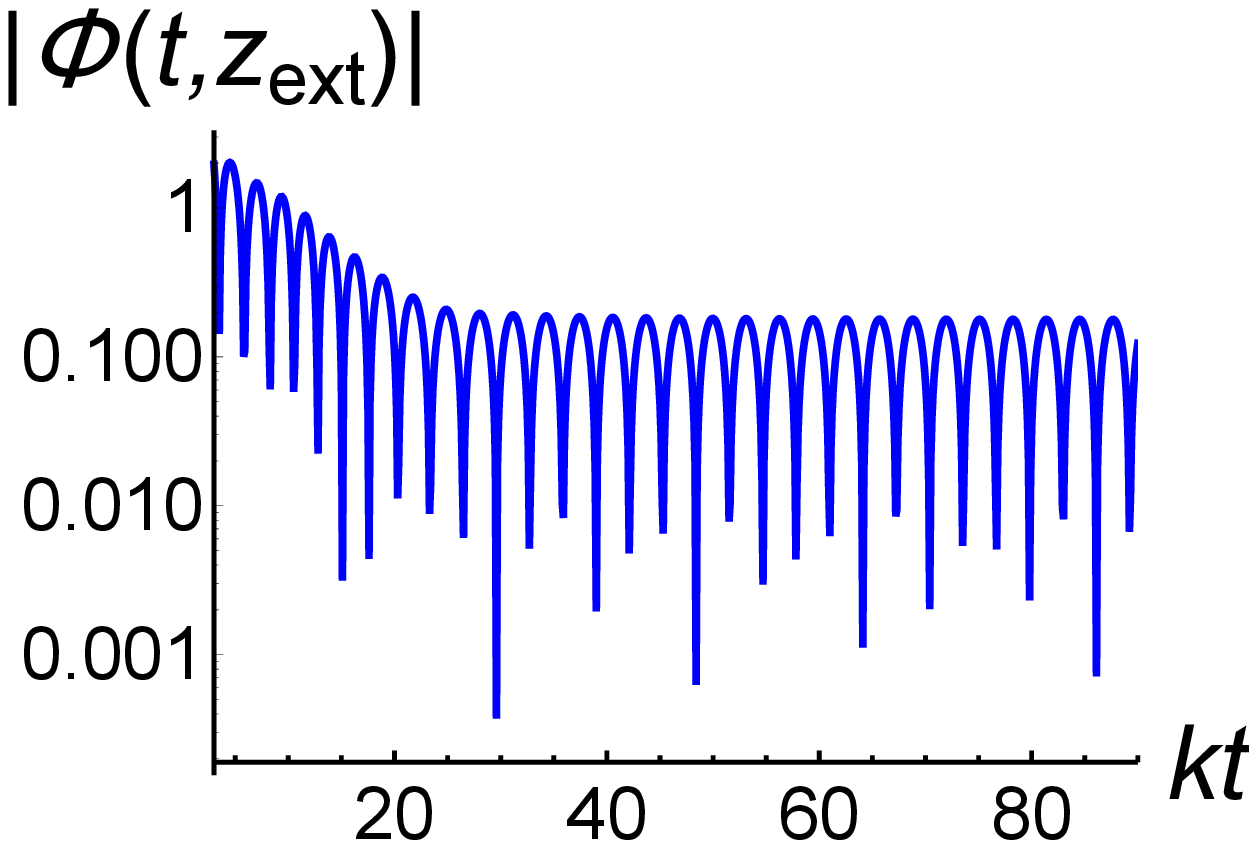}}
		\subfigure[~$kz_{\text{ext}}=10$]{\label{figevenpofilezero10}
		\includegraphics[width=0.22\textwidth]{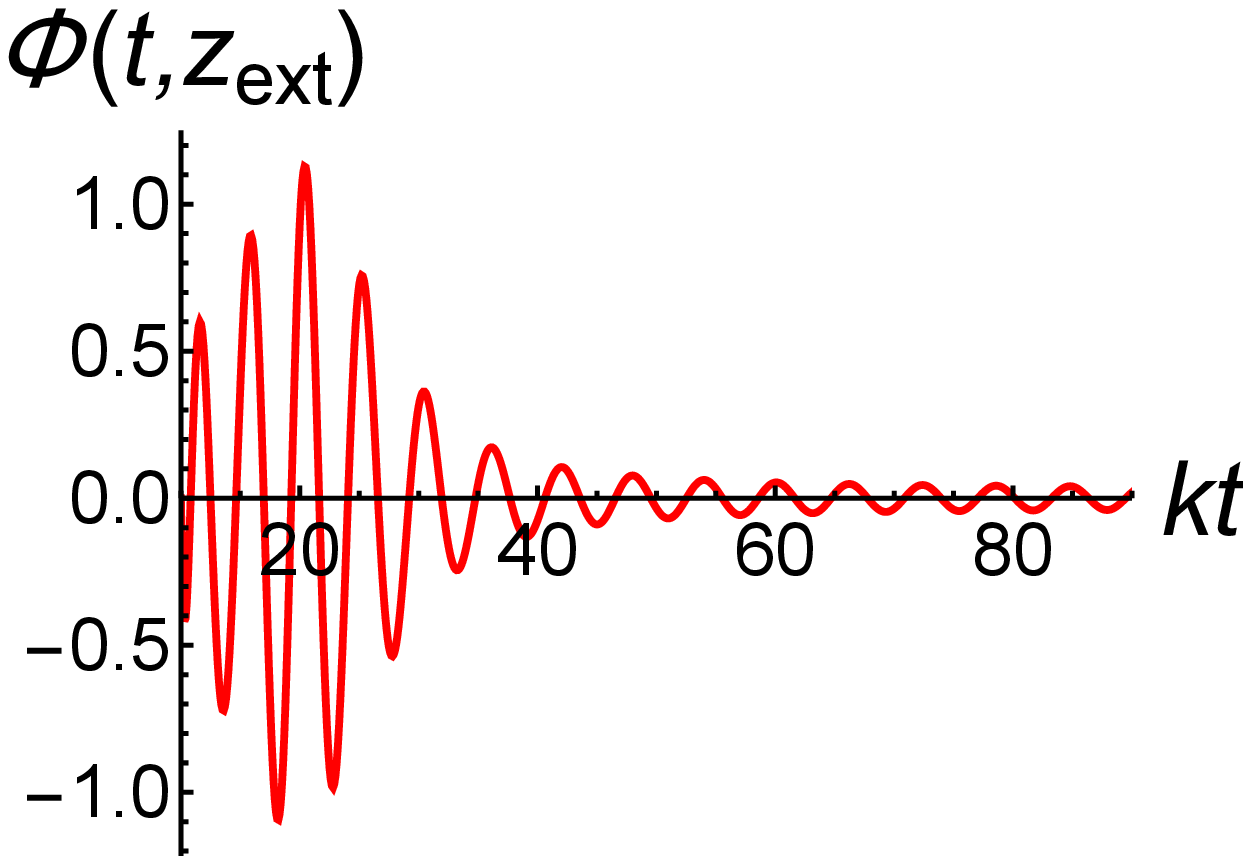}}
	\subfigure[~$kz_{\text{ext}}=10$]{\label{figlogevenpofilezero10}
		\includegraphics[width=0.22\textwidth]{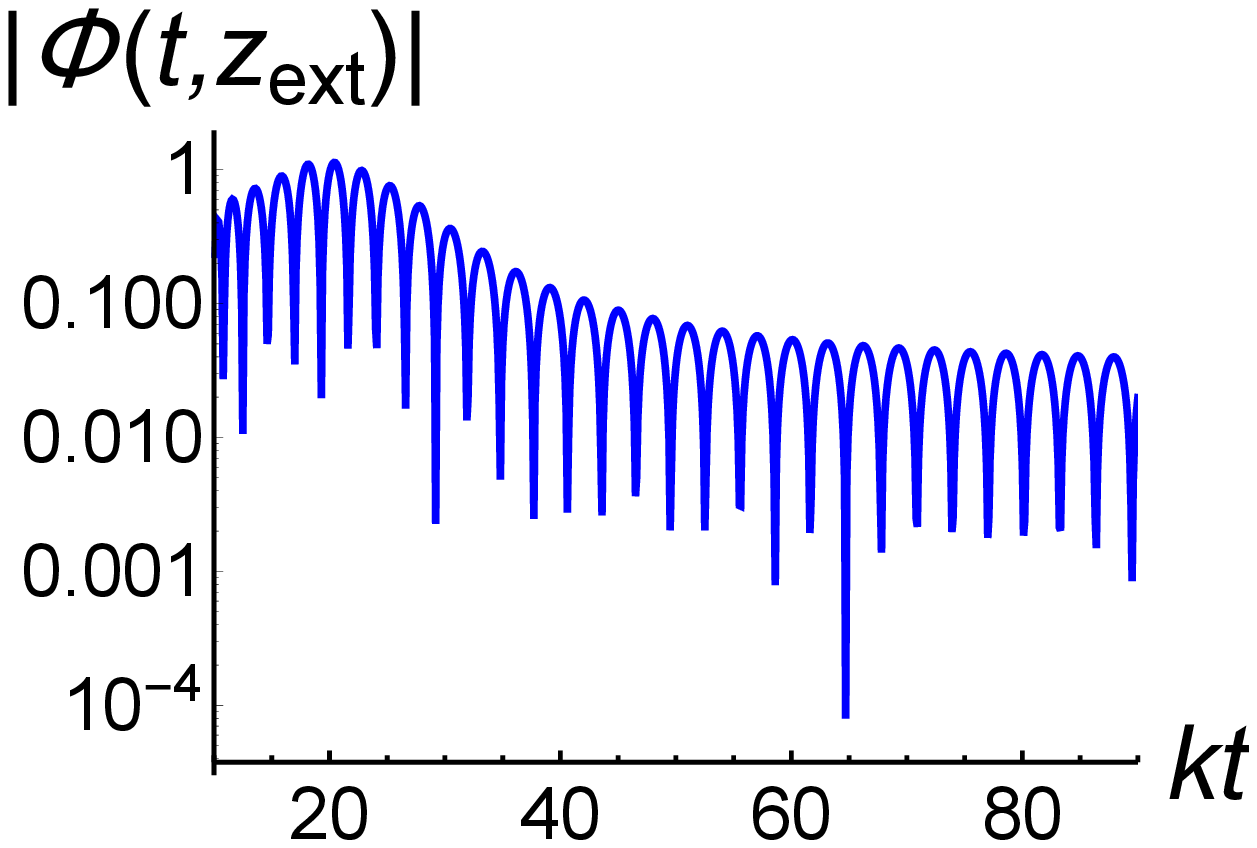}}
	\caption{Left panel: Time evolution of the Gauss pulse at different locations. The signals are extracted at the points $kz_{\text{ext}}=0,~3,~10$. Right panel: Same as left panel but in a logarithmic scale.}\label{extzeromodefig}
\end{figure}

\begin{figure}[htbp]
	\centering
	\includegraphics[width=0.35\textwidth]{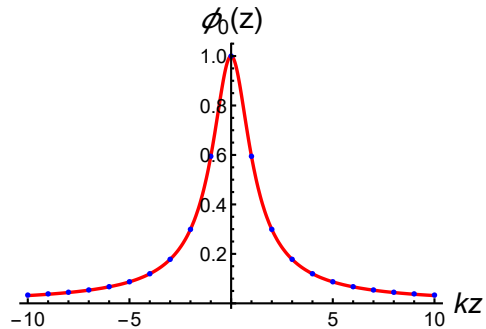}
	\caption{Comparing the results of the zero mode (the blue dots) exited by the Gauss pulse with the analytical zero mode~\eqref{zeromode} (the red curve) obtained from the Schr\"odinger-like equation (\ref{Schrodingerlikeequation}) or equivalently the linear perturbation equation (\ref{conformalequation1}).}\label{zeromodefit}
\end{figure}

Now we consider the numeric evolution of an initial wave packet against the thick brane. We use the $u-v$ coordinate, where $u=t-z$ and $v=t+z$, to perform the evolution of Eq.~\eqref{evolutionequation}. Then Eq.~\eqref{evolutionequation} can be written as
\begin{eqnarray}
	\left(4\frac{\partial^{2}}{\partial u\partial v}+U+a^{2}\right)\Phi=0. \label{uvevolutionequation}
\end{eqnarray}
The incident wave packet is assumed to be a Gaussian pulse,
\begin{eqnarray}
	\Phi(0,v)=e^{\frac{-(v-v_{c})^{2}}{2\sigma^{2}}}, ~~~\Phi(u,0)=e^{\frac{-v_{c}^{2}}{2\sigma^{2}}}.\label{gausspulseinitialwavepacket}
\end{eqnarray}
Here, we focus on the Gaussian pulse with $kv_{c}=5$ and $k\sigma=1$. The parameter $a$ is set to $a/k=1$. $u$ and $v$ belong to $(0,90/k)$. The evolution of the Gauss pulse is shown in Fig.~\ref{extzeromodefig}. In the early time, the waveform is affected by the initial data. Then the waveform evolves into a plane wave. The frequency and the maximum amplitude of the plane wave do not vary with time. From Figs.~\ref{figevenpofilezero0}, \ref{figevenpofilezero3}, \ref{figevenpofilezero10}, we can see that the frequencies of the plane waves do not depend on the extracting points. But the maximum amplitudes of the plane waves depend on the extracting points. Observing the maximum amplitude at each extracting point for the same Gauss pulse, we can see that the final maximum amplitude decreases with $kz_{\text{ext}}$. That is to say, the further away from the brane, the smaller the amplitude. We compare the maximum amplitudes extracted from different points with the profile of the zero mode (\ref{zeromode}). The result is shown in Fig.~\ref{zeromodefit}, which shows that the maximum amplitude as a function of $kz$ is consistent with the analytical zero mode \eqref{zeromode}. Thus, after the pulse hits the brane, the incident pulse excites the zero mode localized on the brane. According to the expression \eqref{decomposition2}, we can obtain the function of the plane wave: $\Phi_{0}(t,z)=e^{-i\omega t}\phi_{0}(z)$. In addition, from the relation $\omega^{2}=m^{2}+a^{2}$, we know that the frequency becomes $\omega=a$ for the zero mode with $m=0$.

\begin{figure}
	\subfigure[~$kz_{ext}=1$]{\label{figoddpofile1}
		\includegraphics[width=0.22\textwidth]{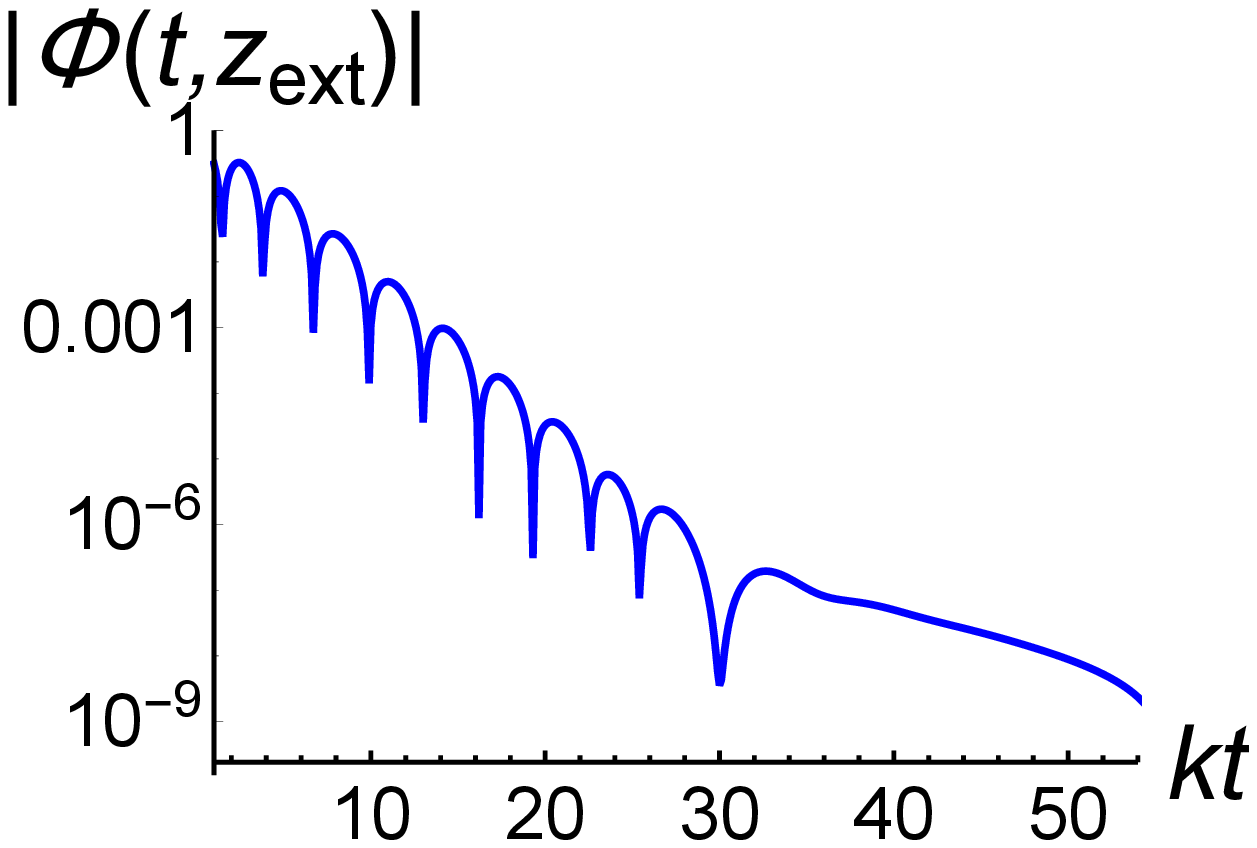}}
	\subfigure[~$kz_{ext}=1$]{\label{figoddpofilea1}
		\includegraphics[width=0.22\textwidth]{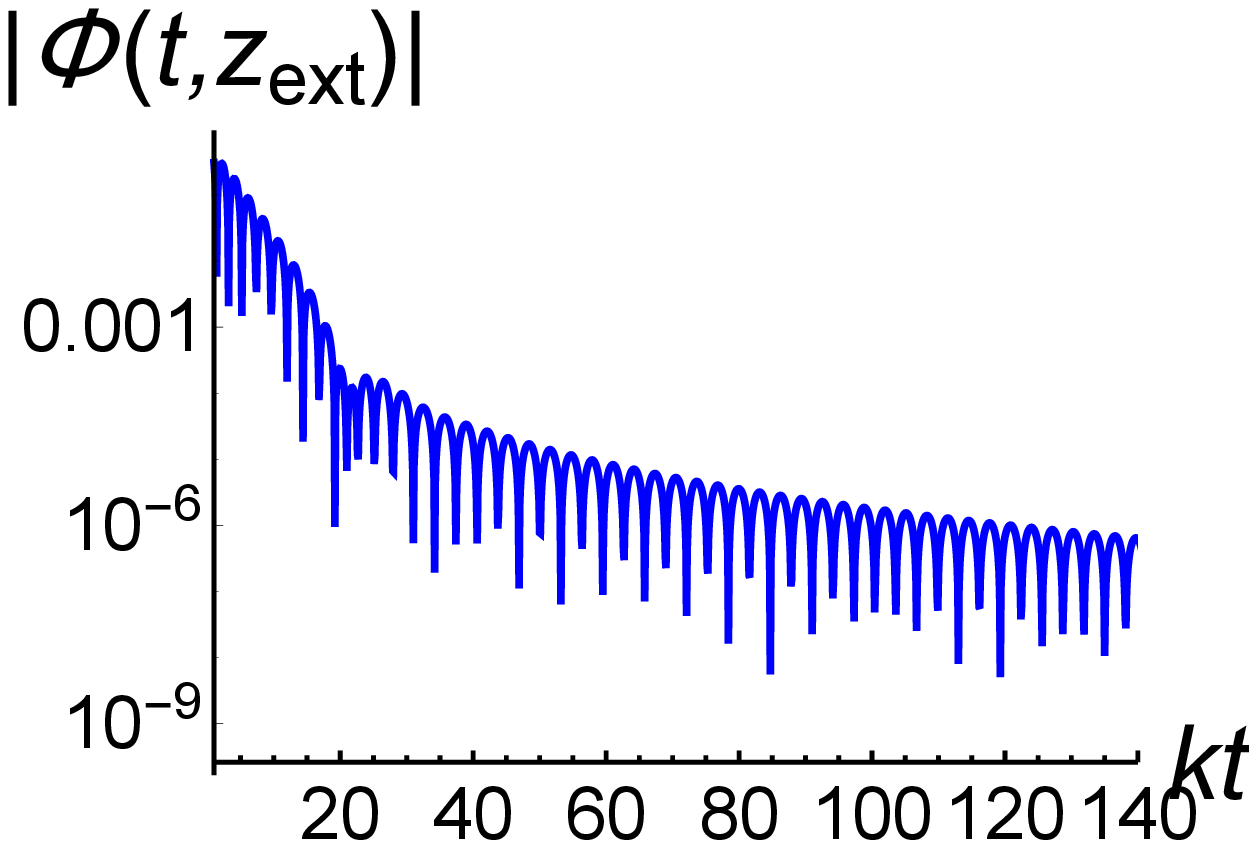}}
	\subfigure[~$kz_{ext}=3$]{\label{figoddpofilez3}
		\includegraphics[width=0.22\textwidth]{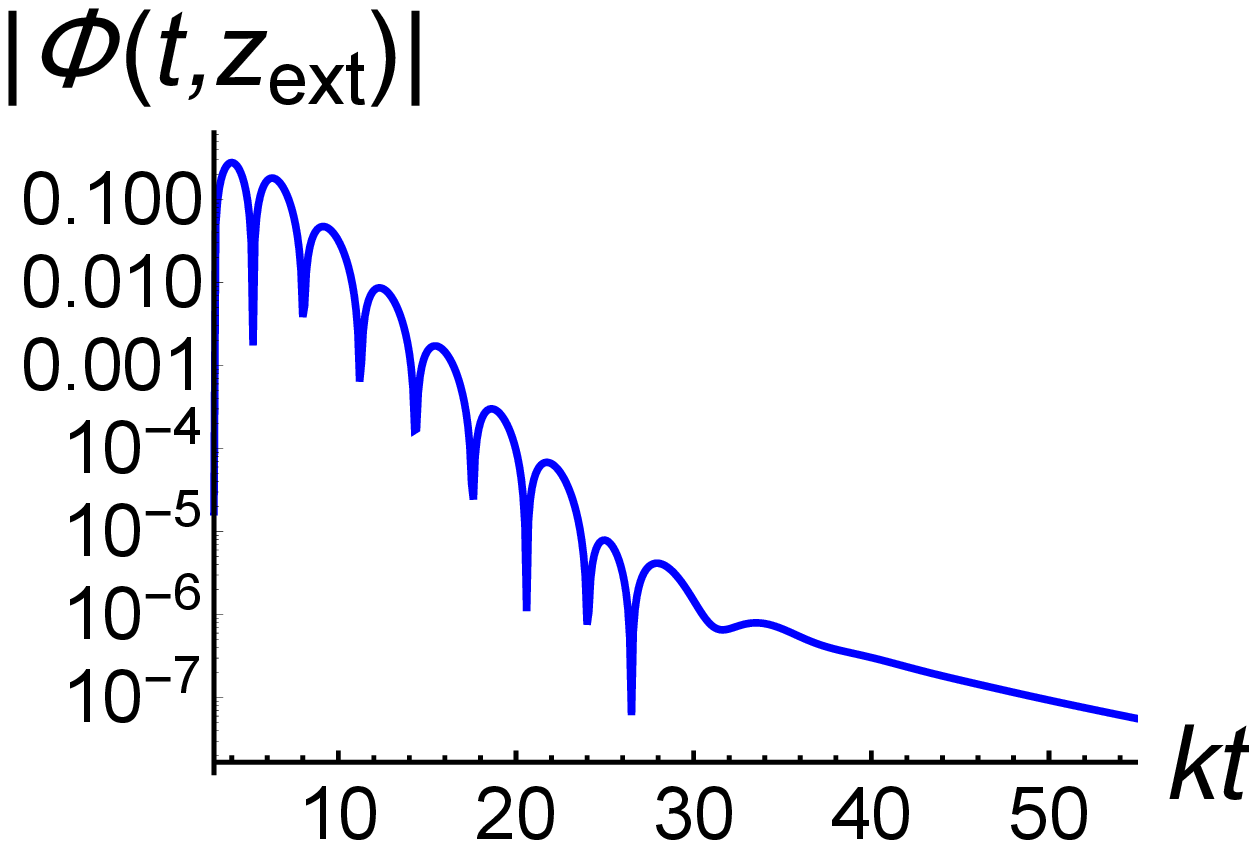}}
	\subfigure[~$kz_{ext}=3$]{\label{figoddpofilea1z3}
		\includegraphics[width=0.22\textwidth]{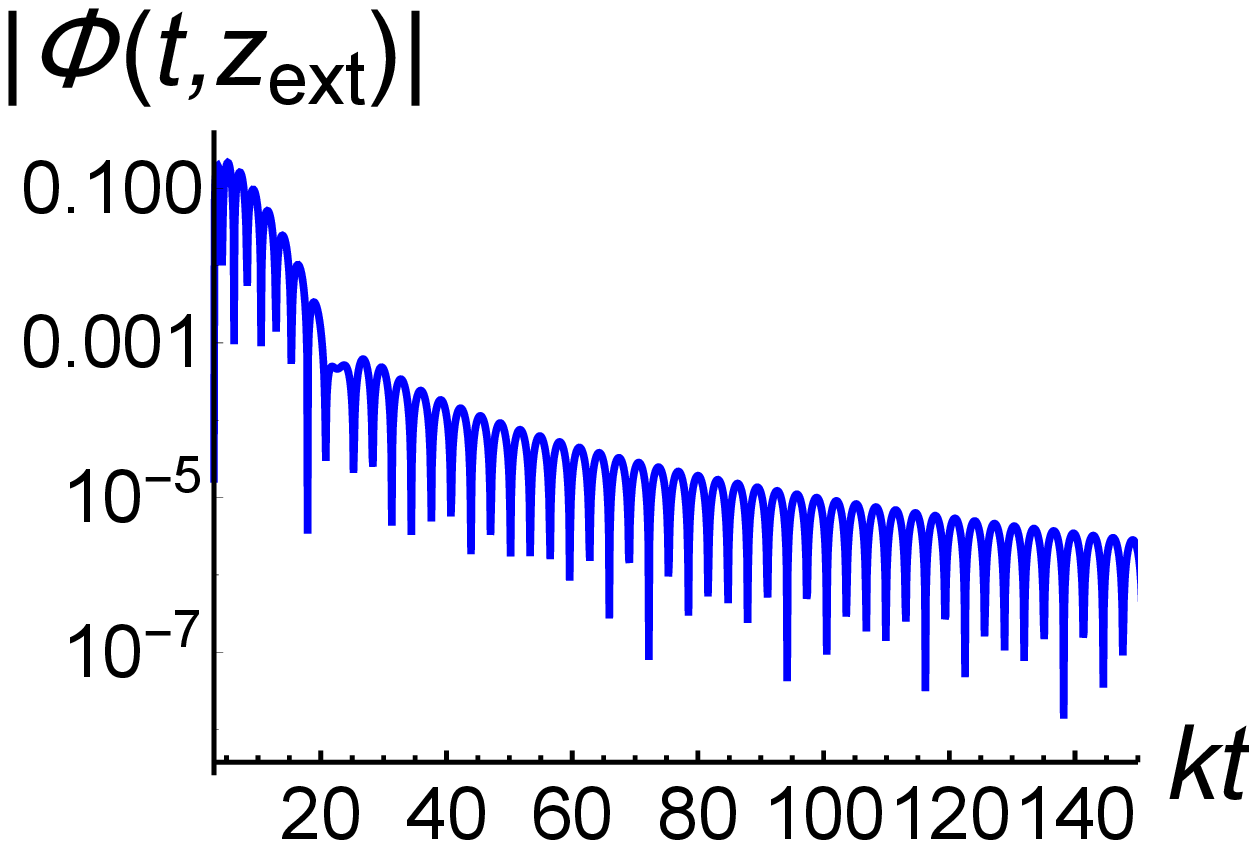}}
	\subfigure[~$kz_{ext}=10$]{\label{figoddpofilez10}
		\includegraphics[width=0.22\textwidth]{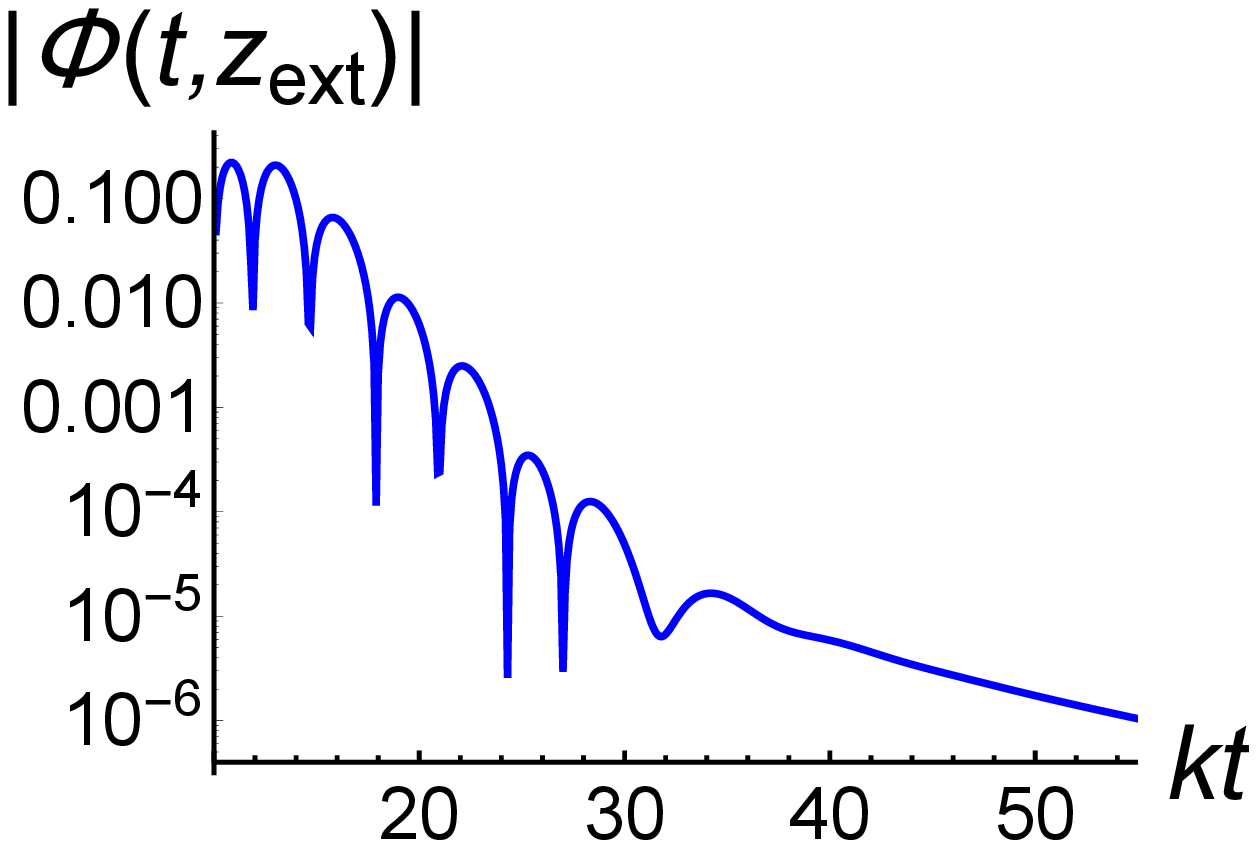}}
	\subfigure[~$kz_{ext}=10$]{\label{figoddpofilea1z10}
		\includegraphics[width=0.22\textwidth]{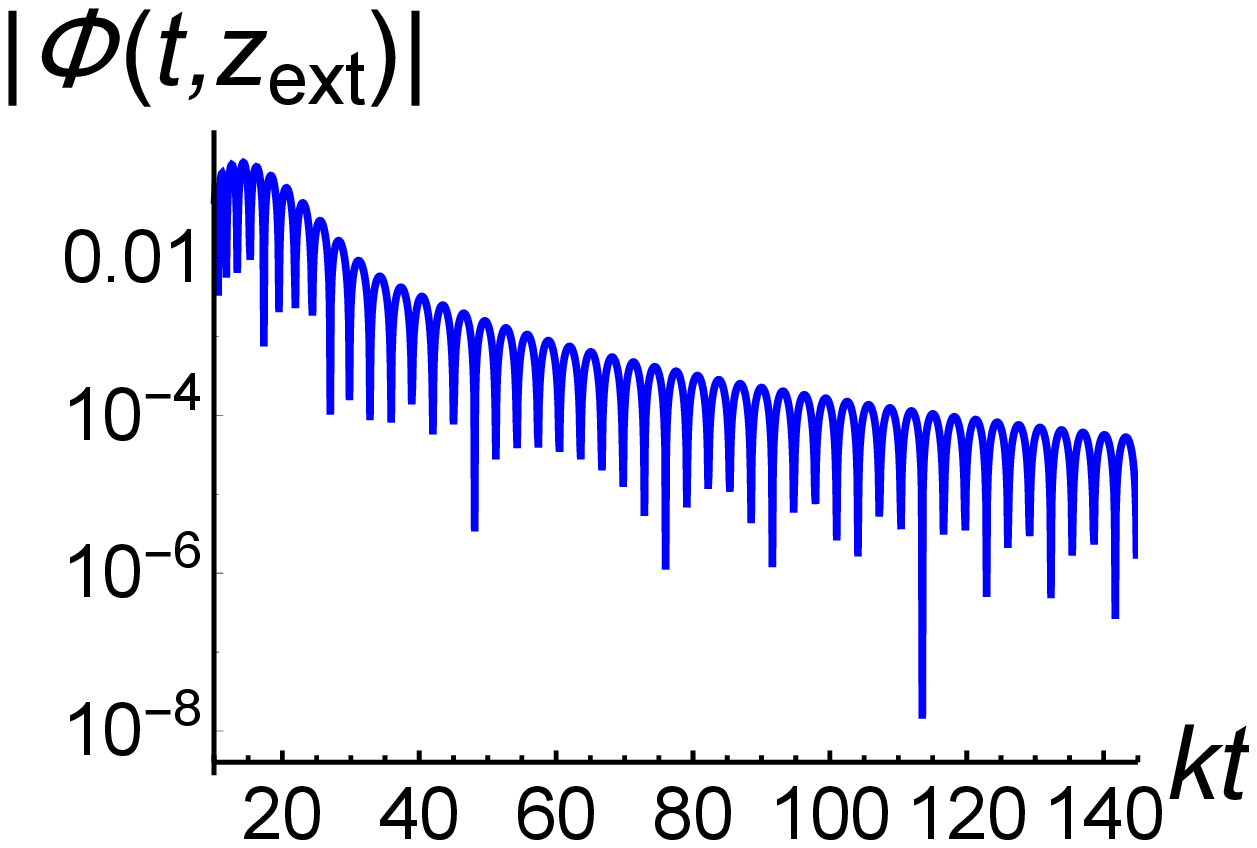}}
	\caption{Left panel: Time evolution of the odd wave packet at different selected extraction points for $a/k=0$. Right panel: Time evolution of the odd wave packet at different selected extraction points for $a/k=1$.}\label{extqnmsfig}
\end{figure}

On the other hand, because the potential is symmetric, the wave functions are either even or odd. Specially, the bound zero mode is even. To investigate the character of the odd QNMs, we give an odd initial wave packet:
\begin{eqnarray}
	\Phi(0,v)=\sin\left(\frac{kv}{2}\right)e^{\frac{-k^2v^{2}}{4}}, \\
	\Phi(u,0)=\sin\left(\frac{ku}{2}\right)e^{\frac{-k^2u^{2}}{4}}.\label{oddinitialwavepacket}
\end{eqnarray}
Plots of the evolution of the waveform are shown in Fig.~\ref{extqnmsfig}. To study the effect of the parameter $a$, we choose $a/k=0$ and $a/k=1$. Obviously, there are two stages through the evolution for the case of $a/k=0$. a) The exponentially decay stage. The frequency and damping time of these oscillations in this stage depend only on the characteristic structure of the thick brane. They are completely independent of the particular initial configuration that causes the excitation of such vibrations. b) The power-law damping stage. This situation is similar to the case of a massless field around a Schwarzschild black hole. Because the first QNM dominates the evolution process, we can obtain the frequency of the first QNM by fitting the evolution data. For the case of Fig.~\ref{figoddpofile1}, the frequency is $\omega/k=1.01079-0.501256i$. This result is good agree with the result of the asymptotic iteration method. For the case of the $a/k=1$, we can see that the quasinormal ringing governs the decay of the perturbation all the time. This is similar to the case of a massive field around a Schwarzschild black hole. It seems that the QNMs in the thick brane model has both two tail characteristics, which is an interesting property. We will investigate the tails of the QNMs for more braneworld models in detail in the future. The above results indicate that there is a normal mode called the zero mode and a series of discrete QNMs in this thick brane model. These modes are the characteristic modes of the brane. The detection of these QNMs can reflect the structure of the brane. From this perspective, these modes are the fingerprints of the brane. This provides a new way for the investigation of the gravitational perturbation in thick brane models.

\begin{figure}
	\subfigure[~ ]{\label{figReomega}
		\includegraphics[width=0.22\textwidth]{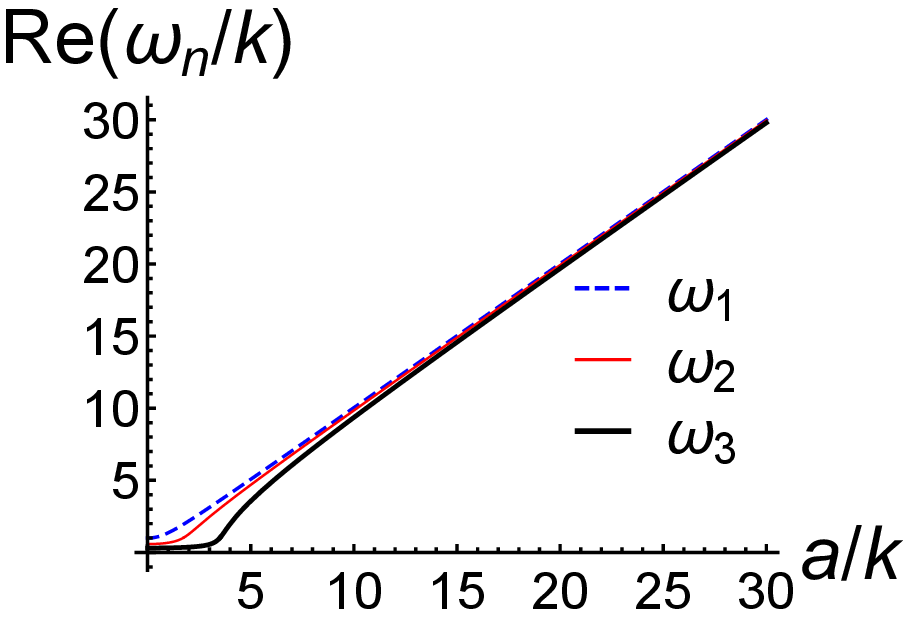}}
	\subfigure[~]{\label{figImomega}
		\includegraphics[width=0.22\textwidth]{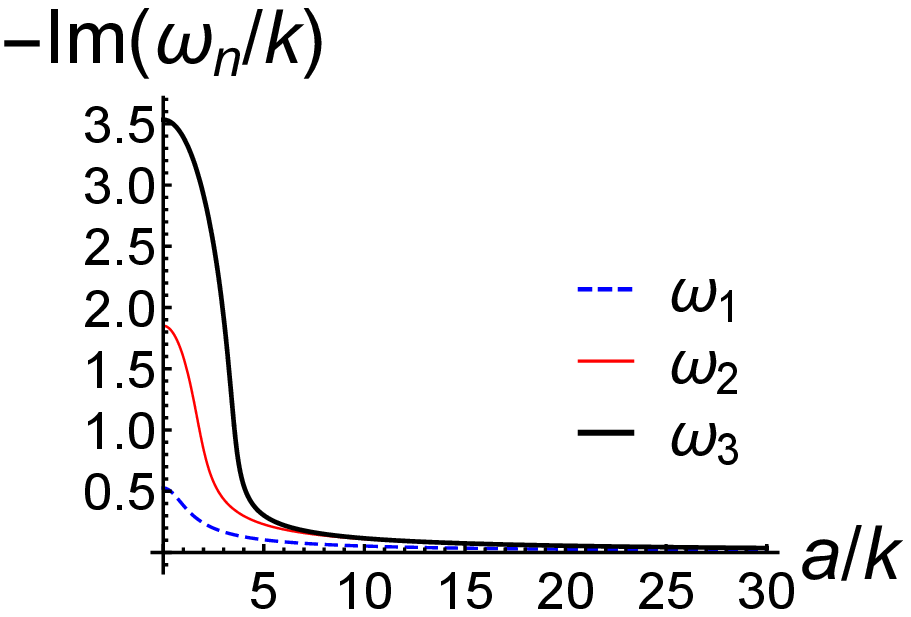}}
	\caption{Left panel: The relations of the real parts of the first three frequencies $\omega_{n}$ and the parameter $a$. Right panel: The relations of the imaginary parts of the first three frequencies $\omega_{n}$ and the parameter $a$.}\label{omegafig}
\end{figure}

To more intuitively understand the character of these modes, following the method of Ref.~\cite{Seahra:2005iq}, we consider a wave packet on the brane
\begin{eqnarray}
\delta h_{\mu\nu}\sim \epsilon_{\mu\nu}\int da \left(\alpha(a)\sum_{n}c_{n}\text{exp}[i(\omega_{n} t-ax)]\right).	
\end{eqnarray}
Here, we consider a motion in the $x$-direction, where $\alpha(a)$ denotes the amplitude of each modes, $c_{n}$ is the expansion coefficient determined by the initial extra dimensional pulse profile, and $n$ runs over the zero mode and QNMs. Obviously, the zero mode acts like it is travelling in a vacuum with the speed of light since $\omega_0=a$. Besides the zero mode, since $\omega_{n}$ has a negative imaginary part, the behavior of each massive mode is that it is propagating in an absorptive medium with a speed  slower than light. If the amplitude $\alpha(a)$ is peaked sharply around some value $a=a_{0}$, then the frequency $\omega_{n}(a)$ can be expanded at that value of $a$. So, we can define the lifetime $\tau_{n}$ and the group velocity~\cite{Jackson:1999cla}
\begin{eqnarray}
	\tau_{n}=\frac{1}{\text{Im}~\omega_{n}},~~~
v_{n}=\text{Re}\left(\partial_{a}\omega_{n}\right)=\text{Re}\left(\frac{a}{\omega_{n}}\right).
\end{eqnarray}
Then we can obtain
\begin{eqnarray}
d_{n}=v_{n}\tau_{n}=\frac{a\text{Re}~\omega_{n}}{\text{Im}~\omega_{n}((\text{Re}~\omega_{n})^{2}+(\text{Im}~\omega_{n})^{2})},
\end{eqnarray}
which is the distance that a massive mode propagates on the brane before its amplitude decreases by a factor of $e$. Since $\omega_{n}=\sqrt{a^{2}+m_{n}^{2}}$, so the real part of $\omega_{n}$ increases with $a$, while the imaginary part $\text{Im}(\omega_{n})$ decreases with $a$. It can also be seen from Fig.~\ref{omegafig}.
This distance is very short for the QNMs with a smaller $a$. For example, when $k=a=10^{-3}$~eV, the distance $d_{n}$ of the first QNM is about $0.2~\text{mm}$. If the distance is of the galactic scale, i.e., $10^{21}~\text{m}$, the frequency of the first QNM is of order $10^{39}~\text{Hz}$ for $k=10^{-3}$~eV. Obviously, it is impossible to find these massive modes from laser interferometer gravitational wave detectors currently in use or under construction~\cite{Bian:2021ini}. These results are consistent with thin brane~\cite{Seahra:2005iq}. Furthermore, Ref.~\cite{Seahra:2005iq} pointed out that, these QNMs might play an important role in the early universe. We expect that the stochastic gravitational wave background could carry potentially information of massive KK modes. In addition, other thick brane models might support long-lived QNMs. In the future, we will investigate the properties of these long-lived modes.

\section{Conclusion and discussion}
\label{Conclusion}
In this paper, we investigated the QNMs of the thick brane model by the semi-analytical and numerical methods. The results obtained by these methods are in good agreement with each other. It shows that there is a zero mode (normal mode) and a series of discrete QNMs in the thick brane model. This is consistent with the results of the RS-II brane~\cite{Seahra:2005iq}. As the characteristic modes of the thick brane, these QNMs play an indispensable role on understanding the structure of the thick brane. This is a new way for the investigation of the gravitational perturbation in thick brane models. It also may provide new ideas for studying thick brane models.

Starting from the solution and the linear metric tensor perturbation given in Ref.~\cite{Gremm:1999pj}, we obtained the wave equation~(\ref{evolutionequation}) and the Schr\"odinger-like equation~(\ref{Schrodingerlikeequation}). Since the Schr\"odinger-like equation can be factorized as a super-symmetric form, we can obtain the super-symmetric partner potential which provides the same spectrum of QNMs of the brane. The super-symmetric partner potential is similar to the effective potentials in the case of the Schwarzschild black hole. Some semi-analytical methods can be used to solve the QNMs. In this way, the QNMs of the thick brane were obtained indirectly. We used the asymptotic iteration method and the WKB approximation to solve the QNMs. The results of the two methods agree with each other in the low overtone region, which can be seen from Table \ref{tab1}. To further confirm the above results, we studied the numerical evolution of the wave equation~(\ref{evolutionequation}). The results show that a zero mode is excited by the incident Gaussian pulse. And the evolution of the odd wave packet reveals the property of the QNMs, which can be seen from Fig.~\ref{extqnmsfig}. In addition, the frequency extracted from the data is consistent with the frequency of the first QNM obtained using the asymptotic iteration method and the WKB approximation. This enhances the credibility of our results. Finally, we investigated the propagation distance $d_{n}$ of the massive mode on the brane. We found that, for the same $m_{n}$, the distance $d_{n}$ increases with the parameter $a$. If the propagation distance is of the galactic scale, the frequency of the massive mode is extremely high, far beyond the ability of the current detectors. However, the massive mode might play a key role in the early universe. It might be detected as a  stochastic gravitational wave background.

Our work could be strengthened in a number of ways. First,we need to develop more methods to calculate higher overtone modes and compare with the asymptotic iteration method. Second, some thick brane models might support long-lived QNMs, which deserve further study. Third, the QNMs of other test fields could be investigated in the future.

\section*{Acknowledgements}
We are thankful to J.~Chen, C.-C.~Zhu for useful discussions. This work was supported by the National Key Research and Development Program of China (Grant No. 2020YFC2201503), the National Natural Science Foundation of China (Grants No.~11875151, No. 12147166, and No.~12047501), the 111 Project under (Grant No. B20063), the China Postdoctoral Science Foundation (Grant No. 2021M701529), and ``Lanzhou City's scientific research funding subsidy to Lanzhou University".


\begin{thebibliography}{99}

\bibitem{Berti:2009kk}
E.~Berti, V.~Cardoso, and A.~O.~Starinets,
\emph{{Quasinormal modes of black holes and black branes}},
 {\emph{Class. Quant. Grav.} {\bfseries 26},
 163001 (2009)},
[{{\ttfamily arXiv:0905.2975}}].


\bibitem{Kokkotas:1999bd}
K.~D.~Kokkotas and B.~G.~Schmidt,
\emph{{Quasinormal modes of stars and black holes}},
 {\emph{Living Rev. Rel.} {\bfseries 2}, 2 (1999)},
[{{\ttfamily  arXiv:gr-qc/9909058}}].


\bibitem{Nollert:1999ji}
H.~P.~Nollert,
\emph{{TOPICAL REVIEW: Quasinormal modes: the characteristic `sound' of black holes and neutron stars}},
 {\emph{Class. Quant. Grav.} {\bfseries 16}, R159 (1999)}.


\bibitem{Konoplya:2011qq}
R.~A.~Konoplya and A.~Zhidenko,
\emph{{Quasinormal modes of black holes: From astrophysics to string theory}},
{\emph{ Rev. Mod. Phys.} {\bfseries 83}, 793 (2011)},
[{{\ttfamily  arXiv:1102.4014}}].


\bibitem{Cardoso:2016rao}
V.~Cardoso, E.~Franzin, and P.~Pani,
\emph{{Is the gravitational-wave ringdown a probe of the event horizon?}}
{\emph{ Phys. Rev. Lett.} {\bfseries 116}, 171101  (2016)},
[erratum: \emph{ Phys. Rev. Lett.}  {\bfseries 117} , 089902 (2016)]
[{{\ttfamily  arXiv:1602.07309}}].



\bibitem{Jusufi:2020odz}
K.~Jusufi, M.~Azreg-A\"\i{}nou, M.~Jamil, S.-W.~Wei, Q.~Wu, and A.-Z.~Wang,
\emph{{Quasinormal modes, quasiperiodic oscillations, and the shadow of rotating regular black holes in nonminimally coupled Einstein-Yang-Mills theory}},
{\emph{ Phys. Rev. D}  {\bfseries 103},  024013 (2021)},
[{{\ttfamily  arXiv:2008.08450}}].


\bibitem{Cheung:2021bol}
M.~H.~Y.~Cheung, K.~Destounis, R.~P.~Macedo, E.~Berti, and V.~Cardoso,
\emph{{Destabilizing the Fundamental Mode of Black Holes: The Elephant and the Flea}},
{\emph{ Phys. Rev. Lett.}  {\bfseries 128}, 111103  (2022)},
[{{\ttfamily arXiv:2111.05415}}].


\bibitem{LIGOScientific:2016aoc}
B.~P.~Abbott \textit{et al.} [LIGO Scientific and Virgo],
\emph{{Observation of Gravitational Waves from a Binary Black Hole Merger}},
{\emph{ Phys. Rev. Lett.}  {\bfseries 116},  061102 (2016)},
[{{\ttfamily arXiv:1602.03837}}].



\bibitem{Kristensen:2015qq}
P.~T.~Kristensen, R.-C.~Ge, and S.~Hughes,
\emph{{Normalization of quasinormal modes in leaky optical cavities and plasmonic resonators}},
{\emph{ Physical Review A},
 {\bfseries 92},
 053810 (2015)},
[{{\ttfamily  arXiv:1501.05938}}].

\bibitem{Randall:1999ee}
L.~Randall and R.~Sundrum, \emph{{A Large mass hierarchy from a small extra dimension}},
 {\emph{Phys. Rev. Lett.}
  {\bfseries 83},  3370 (1999)},
  [{{\ttfamily arXiv:hep-ph/9905221}}].

  \bibitem{Randall:1999vf}
L.~Randall and R.~Sundrum, \emph{{An Alternative to compactification}},
 {\emph{Phys. Rev. Lett.}
  {\bfseries 83},  4690 (1999)},
  [{{\ttfamily arXiv:hep-th/9906064}}].


\bibitem{Shiromizu:1999wj}
T.~Shiromizu, K.~Maeda, and M.~Sasaki,
 {\emph{The Einstein equation on the 3-brane world}},
 {\emph{Phys. Rev. D} {\bfseries 62},  024012 (2000)},
  [{{\ttfamily arXiv:gr-qc/9910076}}].


\bibitem{Tanaka:2002rb}
T.~Tanaka,
{\emph{Classical black hole evaporation in Randall-Sundrum infinite brane world}},
 {\emph{Prog. Theor. Phys. Suppl.} {\bfseries 148},  307 (2003)},
[{{\ttfamily arXiv:gr-qc/0203082}}].

\bibitem{Gregory:2008rf}
R.~Gregory,
 {\emph{Braneworld black holes}},
 {\emph{Lect. Notes Phys.}  {\bfseries 769},  259 (2009)},
[{{\ttfamily arXiv:0804.2595}}].


\bibitem{Jaman:2018ucm}
N.~Jaman and K.~Myrzakulov,
 {\emph{Braneworld inflation with an effective $\alpha$-attractor potential}},
 {\emph{Phys. Rev. D} {\bfseries 99}, 103523 (2019)},
[{{\ttfamily arXiv:1807.07443}}].


\bibitem{Adhikari:2020xcg}
R.~Adhikari, M.~R.~Gangopadhyay, and Yogesh,
 {\emph{Power Law Plateau Inflation Potential In The RS $II$ Braneworld Evading Swampland Conjecture}},
 {\emph{Eur. Phys. J. C} {\bfseries 80}, 899 (2020)},
[{{\ttfamily arXiv:2002.07061}}].

\bibitem{Bhattacharya:2021jrn}
A.~Bhattacharya, A.~Bhattacharyya, P.~Nandy, and A.~K.~Patra,
 {\emph{Islands and complexity of eternal black hole and radiation subsystems for a doubly holographic model}},
 {\emph{JHEP} {\bfseries 05}, 135 (2021)},
[{{\ttfamily arXiv:2103.15852}}].


\bibitem{Akama:1982jy}
K.~Akama,
\emph{{An Early Proposal of `Brane World'}},
{\emph{Lect. Notes Phys.} {\bfseries 176}, 267 (1982)},
[{{\ttfamily arXiv:hep-th/0001113}}].


\bibitem{Rubakov:1983bb}
V.~A.~Rubakov and M.~E.~Shaposhnikov,
\emph{{Do We Live Inside a Domain Wall?}}
{\emph{Phys. Lett. B} {\bfseries 125}, 136 (1983)}.

\bibitem{DeWolfe:1999cp}
O.~DeWolfe, D.~Z.~Freedman, S.~S.~Gubser, and A.~Karch,
\emph{{Modeling the fifth-dimension with scalars and gravity}},
{\emph{Phys. Rev. D} {\bfseries 62}, 046008 (2000)},
[{{\ttfamily arXiv:hep-th/9909134}}].


\bibitem{Gremm:1999pj}
M.~Gremm,
\emph{{Four-dimensional gravity on a thick domain wall}},
{\emph{Phys. Lett. B} {\bfseries 478}, 434 (2000)},
[{{\ttfamily arXiv:hep-th/9912060}}].


\bibitem{Csaki:2000fc}
C.~Csaki, J.~Erlich, T.~J.~Hollowood, and Y.~Shirman,
\emph{{Universal aspects of gravity localized on thick branes}},
{\emph{Nucl. Phys. B} {\bfseries 581}, 309 (2000)},
[{{\ttfamily arXiv:hep-th/0001033}}].



\bibitem{Dzhunushaliev:2010fqo}
V.~Dzhunushaliev and V.~Folomeev,
\emph{{Spinor brane}},
{\emph{Gen. Rel. Grav.} {\bfseries 43}, 1253  (2011)},
[{{\ttfamily arXiv:0909.2741}}].



\bibitem{Dzhunushaliev:2011mm}
V.~Dzhunushaliev and V.~Folomeev,
\emph{{Thick brane solutions supported by two spinor fields}},
{\emph{Gen. Rel. Grav.} {\bfseries 44}, 253  (2012)},
[{{\ttfamily arXiv:1104.2733}}].



\bibitem{Geng:2015kvs}
W.-J.~Geng and H.~Lu,
\emph{{Einstein-Vector Gravity, Emerging Gauge Symmetry and de Sitter Bounce}},
{\emph{Phys. Rev. D} {\bfseries 93}, 044035  (2016)},
[{{\ttfamily arXiv:1511.03681}}].



\bibitem{Melfo2006}
A.~Melfo, N.~Pantoja, and J.~D. Tempo, {\it {Fermion localization on thick branes}},  {\em Phys. Rev. D} {\bf 73},  044033 (2006), [{{\tt arXiv:hep-th/0601161}}].

\bibitem{Almeida2009}
C.~A. Almeida, R.~Casana, M.~M. Ferreira, and A.~R. Gomes, {\it {Fermion localization and resonances on two-field thick branes}},  {\em Phys. Rev. D} {\bf 79},  125022 (2009), [{{\tt arXiv:0901.3543}}].

\bibitem{Zhao2010}
Z.-H. Zhao, Y.-X. Liu, and H.-T. Li, {\it {Fermion localization on asymmetric two-field thick branes}},  {\em Class. Quantum Gravity} {\bf 27},  185001 (2010), [{{\tt arXiv:0911.2572}}].

\bibitem{Chumbes2011}
A.~E.~R. Chumbes, A.~E.~O. Vasquez, and M.~B. Hott, {\it {Fermion localization on a split brane}},  {\em Phys. Rev. D} {\bf 83},  105010 (2011), [{{\tt arXiv:1012.1480}}].

\bibitem{Liu2011}
Y.-X. Liu, Y.~Zhong, Z.-H. Zhao, and H.-T. Li, {\it {Domain wall brane in squared curvature gravity}},  {\em J. High Energy Phys.} {\bf 2011}, 135 (2011), [{{\tt arXiv:1104.3188v2}}].

\bibitem{Xie2017}
Q.-Y. Xie, H.~Guo, Z.-H. Zhao, Y.-Z. Du, and Y.-P. Zhang, {\it {Spectrum structure of a fermion on Bloch branes with two scalar-fermion couplings}}, {\em Class. Quantum Gravity} {\bf 34},  055007 (2017), [{{\tt arXiv:1510.03345}}].

\bibitem{Gu2017}
B.-M. Gu, Y.-P. Zhang, H.~Yu, and Y.-X. Liu, {\it {Full linear perturbations and localization of gravity on $f(R, T)$ brane}},  {\em Eur. Phys. J. C} {\bf 77}, 115 (2017), [{{\tt arXiv:1606.07169}}].

\bibitem{ZhongYuan2017}
Y.~Zhong and Y.-X. Liu, {\it {Linearization of a warped $f(R)$ theory in the higher-order frame}},  {\em Phys. Rev. D} {\bf 95}, 104060 (2017),
[{{\tt arXiv:1611.08237}}].

\bibitem{ZhongYuan2017b}
Y.~Zhong, K.~Yang, and Y.-X. Liu, {\it {Linearization of a warped $f(R)$ theory in the higher-order frame II: The equation of motion approach}}, {\em Phys. Rev. D} {\bf 97},  044032 (2017), [{{\tt arXiv:1708.03737}}].

\bibitem{Zhou2018}
X.-N. Zhou, Y.-Z. Du, H.~Yu, and Y.-X. Liu,
{\it {Localization of gravitino field on $f(R)$-thick branes}},
 {\em Sci. China Physics, Mech. Astron.} {\bf 61}, 110411 (2018),
[{{\tt arXiv:1703.10805}}].



\bibitem{Chen:2020zzs}
J.~Chen, W.-D.~Guo, and Y.-X.~Liu,
{\it {Thick branes with inner structure in mimetic $f(R)$ gravity}},
{\em Eur. Phys. J. C} {\bf 81}, 709 (2021),
[{{\tt arXiv:2011.03927}}].



\bibitem{Hendi:2020qkk}
S.~H.~Hendi, N.~Riazi, and S.~N.~Sajadi,
{\it {$Z_2$-symmetric thick brane with a specific warp function}},
{\em Phys. Rev. D} {\bf 102}, 124034 (2020),
[{{\tt arXiv:2011.11093}}].



\bibitem{Xie:2021ayr}
Q.-Y.~Xie, Q.-M.~Fu, T.-T.~Sui, L.~Zhao, and Y.~Zhong,
{\it {First-Order Formalism and Thick Branes in Mimetic Gravity}},
{\em Symmetry} {\bf 13}, 1345 (2021),
[{{\tt arXiv:2102.10251}}].



\bibitem{Moreira:2021uod}
A.~R.~P.~Moreira, F.~C.~E.~Lima, J.~E.~G.~Silva, and C.~A.~S.~Almeida,
{\it {First-order formalism for thick branes in $f(T,{\mathscr {T}})$ gravity}},
 {\em Eur. Phys. J. C}   {\bf 81}, 1081 (2021),
[{{\tt arXiv:2107.04142}}].


\bibitem{Xu:2022ori}
N.~Xu, J.~Chen, Y.-P.~Zhang, and Y.-X.~Liu,
{\it {Multi-kink brane in Gauss-Bonnet gravity}},
[{{\tt arXiv:2201.10282}}].



\bibitem{Silva:2022pfd}
J.~E.~G.~Silva, R.~V.~Maluf, G.~J.~Olmo, and C.~A.~S.~Almeida,
{\it {Braneworlds in $f(Q)$ gravity}},
[{{\tt arXiv:2203.05720}}].


\bibitem{Xu:2022gth}
Y.-Q.~Xu and X.-D.~Zhang,
{\it {Tensor Perturbations and Thick Branes in Higher Dimensional Gauss-Bonnet Gravity}},
[{{\tt arXiv:2203.13401}}].



\bibitem{Mounaix:2002mm}
P.~Mounaix and D.~Langlois,
{\it {Cosmological equations for a thick brane}},
 {\em Phys. Rev. D}  {\bf 65},  103523 (2002),
[{{\tt arXiv:gr-qc/0202089}}].


\bibitem{Ghassemi:2006qk}
S.~Ghassemi, S.~Khakshournia, and R.~Mansouri,
{\it {Generalized Friedmann equations for a finite thick brane}},
 {\em JHEP}  {\bf 08}, 019 (2006),
[{{\tt arXiv:gr-qc/0605094}}].

\bibitem{Wu:2010stv}
S.-F.~Wu, G.-H.~Yang, and P.-M.~Zhang,
{\it {Cosmological equations and Thermodynamics on Apparent Horizon in Thick Braneworld}},
 {\em Gen. Rel. Grav.}  {\bf 42}, 1601 (2010),
[{{\tt arXiv:0710.5394}}].

\bibitem{Chakraborty:2017qve}
S.~Chakraborty, K.~Chakravarti, S.~Bose, and S.~SenGupta,
{\it {Signatures of extra dimensions in gravitational waves from black hole quasinormal modes}},
{\em Phys. Rev. D} {\bf 97}, 104053 (2018),
[{{\tt arXiv:1710.05188}}].


\bibitem{Prasobh:2014zea}
C.~B.~Prasobh and V.~C.~Kuriakose,
{\it {Quasinormal Modes of Lovelock Black Holes}},
{\em Eur. Phys. J. C} {\bf 74}, 3136 (2014),
[{{\tt arXiv:1405.5334}}].

\bibitem{Dey:2020pth}
R.~Dey, S.~Biswas, and S.~Chakraborty,
{\it {Ergoregion instability and echoes for braneworld black holes: Scalar, electromagnetic, and gravitational perturbations}},
{\em Phys. Rev. D} {\bf  103}, 084019 (2021),
[{{\tt arXiv:2010.07966}}].

\bibitem{Hashemi:2019jlt}
S.~S.~Hashemi, M.~Kord Zangeneh, and M.~Faizal,
{\it {Charged scalar quasi-normal modes for higher-dimensional Born\textendash{}Infeld dilatonic black holes with Lifshitz scaling}},
{\em Eur. Phys. J. C} {\bf 80}, 111 (2020),
[{{\tt arXiv:1901.11367}}].


\bibitem{Chen:2016qii}
C.-H.~Chen, H.-T.~Cho, A.~S.~Cornell, and G.~Harmsen,
{\it {Spin-3/2 fields in $D$-dimensional Schwarzschild black hole spacetimes}},
{\em Phys. Rev. D} {\bf 94}, 044052 (2016),
[{{\tt arXiv:1605.05263}}].


\bibitem{Konoplya:2003dd}
R.~A.~Konoplya,
{\it {Gravitational quasinormal radiation of higher dimensional black holes}},
{\em Phys. Rev. D} {\bf  68}, 124017 (2003),
[{{\tt arXiv:hep-th/0309030}}].


\bibitem{Cardoso:2003vt}
V.~Cardoso, J.~P.~S.~Lemos, and S.~Yoshida,
{\it {Quasinormal modes of Schwarzschild black holes in four-dimensions and higher dimensions}},
{\em Phys. Rev. D} {\bf  69}, 044004 (2004),
[{{\tt arXiv:gr-qc/0309112}}].


\bibitem{Seahra:2004fg}
S.~S.~Seahra, C.~Clarkson, and R.~Maartens,
{\it {Detecting extra dimensions with gravity wave spectroscopy: the black string brane-world}},
{\em Phys. Rev. Lett.} {\bf  94}, 121302 (2005),
[{{\tt arXiv:gr-qc/0408032}}].


\bibitem{Seahra:2006tm}
S.~S.~Seahra,
{\it {Gravitational waves and cosmological braneworlds: A Characteristic evolution scheme}},
{\em Phys. Rev. D} {\bf 74}, 044010  (2006),
[{{\tt arXiv:hep-th/0602194}}].


\bibitem{Chung:2015mna}
H.~Chung, L.~Randall, M.~J.~Rodriguez, and O.~Varela,
{\it {Quasinormal ringing on the brane}},
{\em Class. Quant. Grav.} {\bf  33},  245013 (2016),
[{{\tt  arXiv:1508.02611}}].

\bibitem{Dey:2020lhq}
R.~Dey, S.~Chakraborty, and N.~Afshordi,
{\it {Echoes from braneworld black holes}},
{\em Phys. Rev. D} {\bf   101},  104014  (2020),
[{{\tt  arXiv:2001.01301}}].

\bibitem{Banerjee:2021aln}
I.~Banerjee, S.~Chakraborty, and S.~SenGupta,
{\it {Looking for extra dimensions in the observed quasi-periodic oscillations of black holes}},
{\em JCAP } {\bf 09}, 037 (2021),
[{{\tt arXiv:2105.06636}}].

\bibitem{Mishra:2021waw}
A.~K.~Mishra, A.~Ghosh, and S.~Chakraborty,
{\it {Constraining extra dimensions using observations of black hole quasi-normal modes}},
[{{\tt arXiv:2106.05558}}].


\bibitem{Lin:2022hus}
Z.-C.~Lin, H.~Yu, and Y.-X.~Liu,
{\it {Shortcut in codimension-2 brane cosmology in light of GW170817}},
[{{\tt arXiv:2202.04866}}].


\bibitem{Seahra:2005wk}
S.~S.~Seahra,
\emph{Ringing the Randall-Sundrum braneworld: Metastable gravity wave bound states},
{\emph{Phys. Rev. D} {\bfseries  72}, 066002 (2005)},
[{{\ttfamily arXiv:hep-th/0501175}}].


\bibitem{Seahra:2005iq}
S.~S.~Seahra,
\emph{Metastable massive gravitons from an infinite extra dimension},
{\emph{Int. J. Mod. Phys. D} {\bfseries  14},  2279 (2005)},
[{{\ttfamily arXiv:hep-th/0505196}}].


\bibitem{Tan:2022uex}
Q.~Tan, Y.-P.~Zhang, W.-D.~Guo, J.~Chen, C.-C.~Zhu, and Y.-X.~Liu,
\emph{Evolution of scalar field resonances on braneworld},
[{{\ttfamily arXiv:2203.00277}}].


\bibitem{Clarkson:2005mg}
C.~Clarkson and S.~S.~Seahra,
\emph{Braneworld resonances},
{\emph{Class. Quant. Grav.} {\bfseries  22}, 3653 (2005)},
[{{\ttfamily arXiv:gr-qc/0505145 }}].


\bibitem{Konoplya:2003ii}
R.~A.~Konoplya,
\emph{Quasinormal behavior of the d-dimensional Schwarzschild black hole and higher order WKB approach},
{\emph{Phys. Rev. D} {\bfseries  68}, 024018 (2003)},
[{{\ttfamily arXiv:gr-qc/0303052}}].


\bibitem{Cooper:1994eh}
F.~Cooper, A.~Khare, and U.~Sukhatme,
\emph{{Supersymmetry and quantum mechanics}},
{\emph{Phys. Rept. }{\bfseries 251},  267 (1995)},
[{{\ttfamily arXiv:hep-th/9405029}}].



\bibitem{Ge:2018vjq}
B.-X.~Ge, J.~Jiang, B.~Wang, H.-B.~Zhang, and Z.~Zhong,
\emph{Strong cosmic censorship for the massless Dirac field in the Reissner-Nordstrom-de Sitter spacetime},
{\emph{JHEP} {\bfseries  01}, 123 (2019)},
[{{\ttfamily arXiv:1810.12128}}].

\bibitem{Ciftci:2003As}
H. Ciftci, R. L. Hall, and N. Saad,
\emph{Asymptotic iteration method for eigenvalue problems},
{\emph{Journal of Physics A},  {\bfseries 36},  11807 (2003)},
[{{\ttfamily arXiv:math-ph/0309066}}].


\bibitem{ciftci:2005co}
H. Ciftci, R. L. Hall, and N. Saad,
\emph{Construction of exact solutions to eigenvalue problems by the asymptotic iteration method},
{\emph{Journal of Physics A: Mathematical and General},
 {\bfseries 38}, 1147 (2005)},
[{{\ttfamily arXiv:math-ph/0412030}}].



\bibitem{Cho:2011sf}
H.-T.~Cho, A.~S.~Cornell, J.~Doukas, T.-R.~Huang, and W.~Naylor,
\emph{A New Approach to Black Hole Quasinormal Modes: A Review of the Asymptotic Iteration Method},
{\emph{Adv. Math. Phys.} {\bfseries  2012}, 281705 (2012)},
[{{\ttfamily arXiv:1111.5024}}].

\bibitem{Konoplya:2019hlu}
R.~A.~Konoplya, A.~Zhidenko, and A.~F.~Zinhailo,
\emph{Higher order WKB formula for quasinormal modes and grey-body factors: recipes for quick and accurate calculations},
{\emph{Class. Quant. Grav.} {\bfseries  36}, 155002 (2019)},
[{{\ttfamily arXiv:1904.10333}}].



\bibitem{schutz:1985bl}
 B. F. Schutz and C. M. Will,
\emph{Black hole normal modes: a semianalytic approach},
{\emph{The Astrophysical Journal}, {\bfseries  291}, L33 (1985)}.



\bibitem{Iyer:1986np}
S.~Iyer and C.~M.~Will,
\emph{Black Hole Normal Modes: A {WKB} Approach. 1. Foundations and Application of a Higher Order {WKB} Analysis of Potential Barrier Scattering},
{\emph{Phys. Rev. D} {\bfseries 35}, 3621 (1987)}.



\bibitem{Konoplya:2004ip}
R.~A.~Konoplya,
\emph{Quasinormal modes of the Schwarzschild black hole and higher order WKB approach},
{\emph{J. Phys. Stud.} {\bfseries  8}, 93 (2004).}


\bibitem{Jackson:1999cla}
J. D. Jackson,
\emph{Classical Electrodynamics, 3rd edition},
{\emph{Wiley, New York}, 1999}.



\bibitem{Bian:2021ini}
L.~Bian, R.-G.~Cai, S.~Cao, Z.~Cao, H.~Gao, Z.-K.~Guo, K.~Lee, D.~Li, J.~Liu, and Y.~Lu, \textit{et al.},
\emph{The Gravitational-wave physics II: Progress},
{\emph{Sci. China Phys. Mech. Astron.} {\bfseries  64}, 120401 (2021)},
[{{\ttfamily arXiv:2106.10235}}].


\end{thebibliography}
\end{document}